\documentclass[aps,prb,10pt,twocolumn,amsmath,amssymb,floatfix,superscriptaddress]{revtex4-1}

\usepackage{graphicx}
\usepackage{txfonts}
\usepackage{braket}
\usepackage[colorlinks,citecolor=blue,linkcolor=blue,urlcolor=blue,plainpages=false,pdfpagelabels]{hyperref}

\usepackage{filecontents}

\begin{document}

	\title{Moving Majorana bound states between distinct helical edges across a quantum point contact}

	\author{Alessio Calzona}
	\affiliation{Institute of Theoretical Physics and Astrophysics,
	University of W\"urzburg, 97074 W\"urzburg, Germany}
	\author{Bj\"orn Trauzettel}
	\affiliation{Institute of Theoretical Physics and Astrophysics,
	University of W\"urzburg, 97074 W\"urzburg, Germany}
	\affiliation{W\"urzburg-Dresden Cluster of Excellence ct.qmat, Germany}
	
	\date{\today}

	\begin{abstract}
		Majorana bound states are zero-energy excitations of topological superconductors which obey non-Abelian exchange statistics and are basic building blocks for topological quantum computation. In order to observe and exploit their extraordinary properties, we need to be able to properly manipulate them, for instance, by braiding a couple of them in real space. We propose a setup based on the helical edges of two-dimensional topological insulators (2DTI) which allows for a high degree of tunability by only controlling a handful of superconducting phases. In particular, our setup allows to move the Majoranas along a single edge as well as to move them across two different edges coupled by a quantum point contact. Robustness against non-optimal control of the phases is also discussed. This proposal constitutes an essential step forward towards realizing 2DTI-based architectures capable of performing braiding of Majoranas in a feasible way.  
	\end{abstract}
	
	\maketitle
	
	\section{Introduction} 
	Topological superconductors have been predicted to support Majorana bound states (MBS), zero energy midgap modes featuring non-Abelian braiding statistics. In addition to the interest in a fundamentally new excitation, the enormous potential of MBS as building blocks for topological quantum computation has made their study one of the most active research fields in condensed matter physics \cite{alicea12,beenakker13,kitaev01,nayak08,Aasen16}. In this respect, the most advanced experimental platform to engineer MBS is represented by semiconducting nanowires with strong spin-orbit coupling, proximitized by a conventional superconductor, in presence of an external magnetic field \cite{Oreg10,Lutchyn10,Higginbotham2015}. This setup has been tested in several ground-breaking experiments, which provided strong evidence for the existence of MBS by measuring zero-bias conductance peaks \cite{mourik12,Albrecht16,Deng16,Nichele2017,Gul2018,Deng2012,Lee2012}.
	
	Several interesting alternative setups have been proposed, relying for example on magnetic adatoms \cite{nadjperge14,Pawlak2016,Feldman17,Ruby15}, vortices in topological superconductors \cite{Roising2019,Xu2015,Beenakker2019}, quantum dots \cite{Malciu2018,Prada2017}, or planar Josephson junctions \cite{Hegde2019,Fu2008,Choi2018,Choi2019,Guiducci2019,Guiducci2018,Pientka2017,Fornieri2019}. In this respect, a promising system which hosts MBS is based on the helical edge states of quantum spin Hall insulators (QSHI) \cite{Konig07,Bruene2012,Knez11,Bernevig06}. In this setup, MBS emerge when parts of the edge are gapped out by superconducting and ferromagnetic barriers \cite{Fu2009,Crepin2014,Crepin2015,Keidel2018,Li2016,Vayrynen2015,Borla2018}. While being potentially more robust against multi-mode and disorder effects, the experimental quest for MBS in QSHI-based devices has proven to be challenging. Nonetheless, superconductivity has been successfully proximity induced into QSHI and experiments based on helical Josephson junction have provided first evidence for the formation of MBS by inspecting (missing) Shapiro steps \cite{Bocquillon2016}. Moreover, recent breakthroughs in the fabrication of QSHI-based devices allow to couple different helical edges through a quantum point contact (QPC) \cite{Strunz2019}.% as well as to locally deposit ferromagnetic islands. 
	
	Helical edge states are therefore likely to become a fertile playground to study the emergence of MBS and their particular braiding properties, whose experimental observation is still lacking. In order to inspect the latter, indeed, systems with a high degree of tunability are required: in general, one has to deal with multiple couples of MBS, to tune their couplings \cite{Park2015}, and/or to vary their positions \cite{Mi2013}. 
	
	Seeking a realistic QSHI-based platform which provides the desired tunability represents the main task of this paper. The starting point is the well-known SFS architecture \cite{Fu2009}, where a single helical edge state is gapped by a finite ferromagnetic region (F) which lies in between two semi-infinite superconducting barriers (S). In this system, a single couple of MBS emerge and its hybridization can be tuned by acting on the superconducting phases \cite{Keidel2018,Crepin2014}. Remarkably, we demonstrate that the insertion of an additional finite superconducting region greatly enhances the versatility of the system, allowing to physically move the MBS along the edge. Such a displacement, which is again controlled exclusively by the superconducting phases, turns out to be independent of the degree of hybridization.
	
	More importantly, the proposed SSFS architecture represents the fundamental building block of a multi-edge setup, where one can move several zero-energy MBS within the whole system (i.e. even between distinct edges). In this paper, we focus in particular on two edges of a 2DTI with SSFS geometry, locally coupled by a QPC. While being experimentally feasible, the proposed setup features great versatility when it comes to the manipulation of Majoranas. In particular, it allows us to move one zero-energy MBS from one edge to the other one by exploiting electron tunneling at the QPC, while the second MBS is kept fixed. Such a straightforward operation, performed by controlling only a handful of superconducting phases, clearly represent an essential step towards physical braiding of Majoranas in 2DTI-based architectures.
	
	\section{Moving Majoranas along a single edge} 
\begin{figure}
		\centering
		\includegraphics[width=1\linewidth]{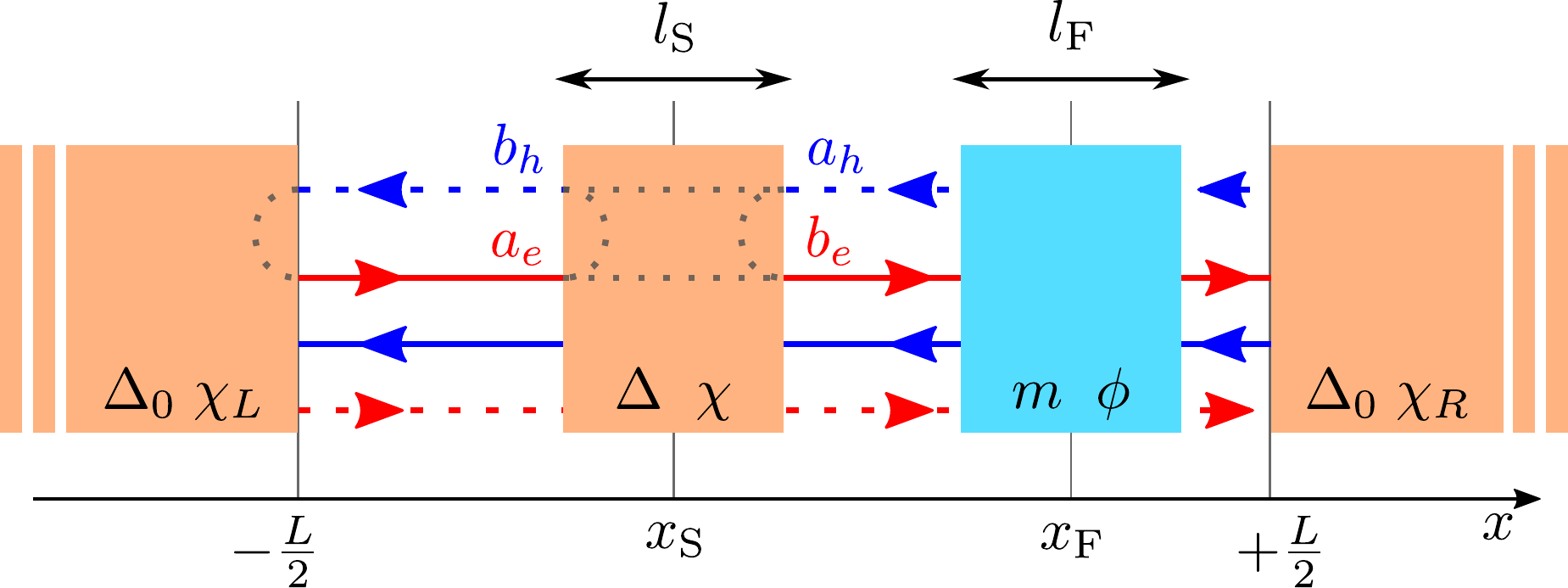}
		\caption{The SSFS setup, a single helical edge gapped by superconductors (orange boxes) and ferromagnet (blue box).
		Electrons (holes) are indicated by solid (dashed) lines, whose color refer to their spin: red (blue) for spin up (down). Some scattering processes, discussed in the main text, are depicted with gray dotted lines. 
		%(B) Energy splitting $\epsilon_{\rm Maj}$ of the two Majoranas as a function of the superconducting phase differences, with $\Delta=m=\Delta_0$. (C-E) Zero-energy Majorana wavefunctions $|\varphi_{0,j}(x)|^2(x)$ along the edges, for different values of the phase difference $\chi-\chi_{\rm L}=0$ (C), $0.9\pi$ (D), $\pi$ (E). The left MBS (purple) moves along the edge while the other one (green) remains fixed. The dotted black lines shows (twice) the zero-energy lDOS $2A_0(x)$. We choose $m=2.5\Delta=2.5\Delta_0$, $l_{\rm S}=l_{\rm F}=\xi=\Delta_0^{-1}$,$L=7\xi$,  $x_{\rm F}=-x_{\rm S}=1.5 \xi$. Wavefunctions and lDOS are plotted in units $\xi^{-1}$. 
		}
		\label{fig:setup}
	\end{figure}

At first, we focus on a single helical edge channel of a QSHI gapped by both superconducting and ferromagnetic regions, as shown in Fig.~\ref{fig:setup} (A). The system Hamiltonian can be conveniently expressed as $H = \tfrac{1}{2} \int dx\;  \Psi^\dagger \mathcal{H}_{\rm BdG} \Psi $ with the well-known Bogoliubov-de Gennes (BdG) Hamiltonian
	\begin{equation}
	\label{eq:BdG}
	\mathcal{H}_{\rm BdG} (x) = p_x \tau_3 \sigma_3 + \vec{m}(x) \cdot \vec{\sigma} + \vec{\Delta}(x) \cdot \vec{\tau}-\mu(x) \, \tau_3.
	\end{equation}
	and the Nambu spinor $\Psi = (\psi_{R\uparrow},\psi_{L\downarrow},\psi_{L\downarrow}^\dagger,-\psi_{R\uparrow}^\dagger)^T$. The electron field operators $\psi_{r,s}$ annihilate a right- ($r=R$) or left- ($r=L$) moving particle with spin $s=\uparrow, \downarrow$ quantized along the $z$ axis. The Pauli matrices $\vec{\sigma}=(\sigma_1,\sigma_2,\sigma_3)$ [ $\vec{\tau}=(\tau_1,\tau_2,\tau_3)$] act on spin (particle-hole) space, $p_x = -i \partial_x$ is the momentum operator, $\mu(x)$ is the chemical potential and we have set both $\hbar = 1$ and the Fermi velocity $v_{\rm F} =1$. The superconducting pairing and the Zeeman coupling reads $\vec{\Delta} = (\Delta \cos\chi, \Delta \sin\chi, 0)^T$ and $\vec{m}=(m \cos\phi, m\sin\phi, m_z)^T $, respectively. 
	As depicted in Fig.\ \ref{fig:setup}, the system we are interested in consists of three normal gapless regions with $m=\Delta=0$ which lie in between regions gapped by either a finite pairing potential $\Delta\neq 0$ (S) or by a finite in-plane magnetization $m\neq0$ (F). The semi-infinite superconductors at the two ends $x=\pm L/2$ lead to perfect Andreev reflections within the superconducting gap $\Delta_0$. For the sake of simplicity, in the following we will consider $\mu=m_z=0$ everywhere. The results we find, however, hold also in presence of finite chemical potential and/or magnetization along the $z$ direction [see Appendix \ref{app:C}]. 	Moreover, we assume all the parameters to be uniform within each region.
	
	It is well known that solutions of the BdG equation $	\mathcal{H}_{\rm BdG} \, \varphi_{\epsilon} =\epsilon  \, \varphi_{\epsilon} $ with energies $|\epsilon|<\Delta_0$ represent mid-gap bound states, described by the Nambu wavefunction $\varphi_{\epsilon}(x)$. The built-in particle-hole symmetry of the BdG Hamiltonian ensures that the bound states always come in pairs with opposite energies. Indeed, given an eigenstate $\varphi_{\epsilon}$, its charge-conjugated partner $\mathcal{C}\varphi_{\epsilon}$ is still an eigenstate of $\mathcal{H}_{\rm BdG}$ with opposite energy $-\epsilon$, where the operator $\mathcal{C}=\mathcal{K} \sigma_2\tau_2$ with $\mathcal{K}$ the complex conjugation \cite{Crepin2014}. As a remarkable consequence, whenever present, bound states at zero energy are always (at least) double degenerate and represent Majorana fermions. Indeed, it is always possible to describe the two states in terms of two wavefunctions  which are invariant under charge-conjugation, i.e. $\mathcal{C}\varphi_{0,j}=\varphi_{0,j}$ with  $j=\alpha,\beta$ \cite{Crepin2014,Chamon10}. In general, the degeneracy can be lifted by acting on the system parameters. Then, the two MBS hybridize and acquire finite excitation energies $\pm \epsilon_{\rm Maj}$. 
	
	In the following, we argue that the SSFS system allows us to selectively move the Majorana wavefunctions by controlling only two superconducting phases. Moreover, such manipulation can be performed without modifying $\epsilon_{\rm Maj}$, in particular while keeping the MBS at zero energy. To properly set the stage for our result, however, it is worth it to briefly review the simpler SFS geometry \cite{Fu2009,Keidel2018,Crepin2014,Crepin2015}. It can be seen as a limiting case of our setup for $l_{S}\to0$. In this setup, the system hosts a couple of zero-energy MBS for $\chi_R-\chi_L = \pi$. By acting on the superconducting phases, their energy splitting can be tuned up to a maximum value $\epsilon_{\rm Maj}\propto e^{-m\, l_{\rm F}}$ reached when $\chi_R-\chi_L = 0$. Unfortunately, the lack of additional knobs does not allow to control the position of the Majorana wavefunctions, which are always localized on the two sides of the ferromagnetic region \cite{Crepin2014}. Such a limitation can be nicely overcome just by considering the presence of an additional superconductor with finite $l_{\rm S}>0$.  
	
\begin{figure}
		\centering
		\includegraphics[width=1\linewidth]{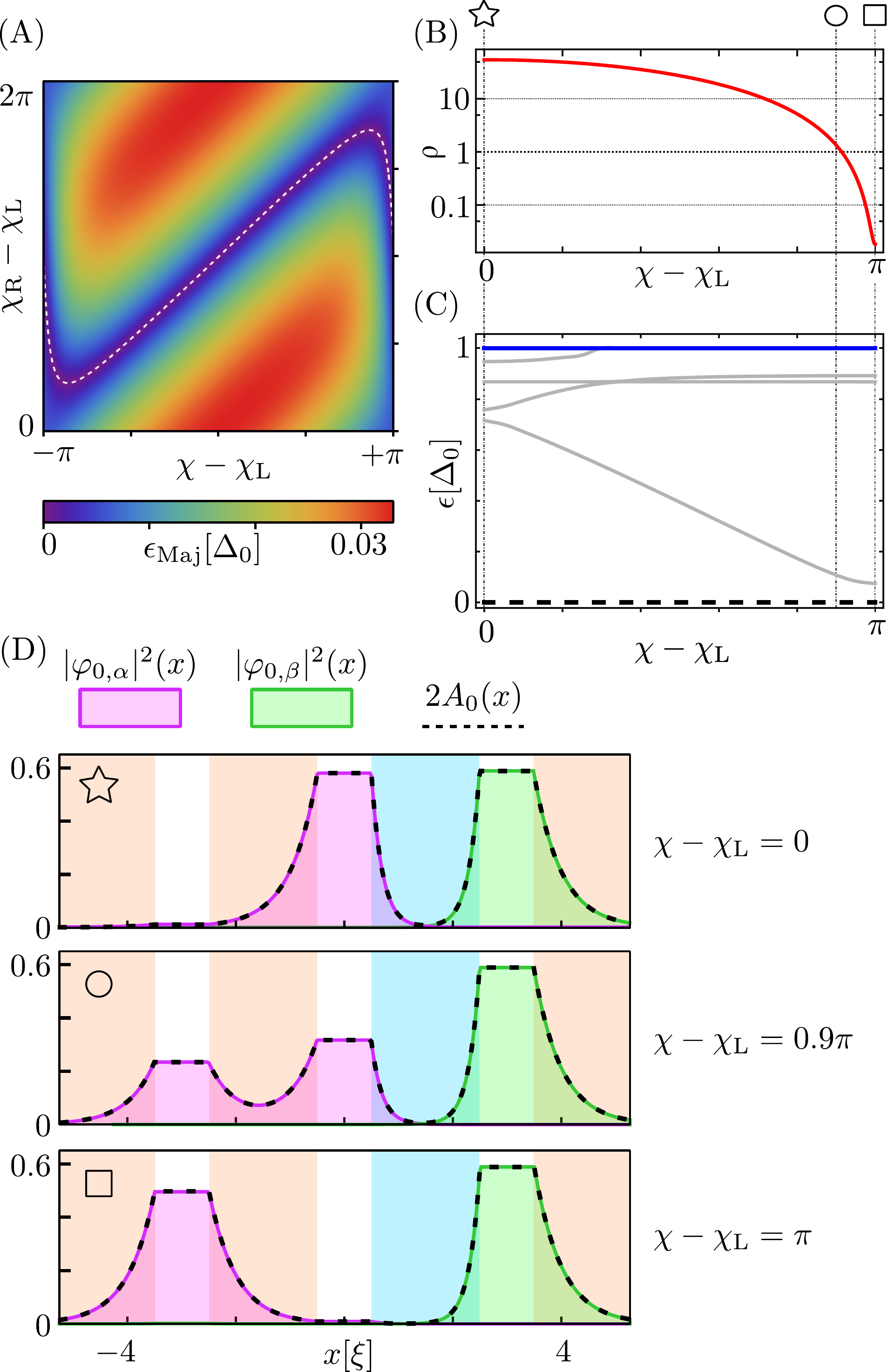}
		\caption{(A) Energy splitting $\epsilon_{\rm Maj}$ of the two Majoranas as a function of the superconducting phase differences, with $\Delta=m=\Delta_0$. (B) Logarithmic plot of the ratio $\rho$ as a function of $\chi-\chi_L$, describing the localization of a zero-energy Majorana to the right ($\rho>1$) or to the left ($\rho<1$) of the central finite superconductor. (C) Spectrum of the system as a function of $\chi-\chi_L$, with $\chi_R$ satisfying the zero-energy condition of Eq.~\eqref{eq:zero_cond}. The zero-energy Majoranas are depicted with the black-dashed line. Other ABS are plotted with gray lines. The superconducting gap $\Delta_0$ is marked in blue. We consider $m=2.5\Delta=2.5\Delta_0$. (D) Zero-energy Majorana wavefunctions $|\varphi_{0,j}(x)|^2(x)$ along the edge, for different values of the phase difference $\chi-\chi_{\rm L}=0$ (star), $0.9\pi$ (circle), $\pi$ (square). The left MBS (purple) moves along the edge while the other one (green) remains fixed. The dotted black lines shows twice the zero-energy lDOS $2A_0(x)$. We choose again $m=2.5\Delta=2.5\Delta_0$. Wavefunctions and lDOS are plotted in units $\xi^{-1}$.  All the plots are computed for the following geometry: $l_{\rm S}=l_{\rm F}=\xi=\Delta_0^{-1}$,$L=7\xi$,  $x_{\rm F}=-x_{\rm S}=1.5 \xi$. 
		}
		\label{fig:fig2}
	\end{figure}

	In order to find the mid-gap bound states of the SSFS system, depicted in Fig.~\ref{fig:setup}, we employ scattering theory. The computations of the scattering matrices, lengthy but straightforward, are described in Appendices \ref{app:A} and \ref{app:B}. For clarity, we only present a specific example which helps to understand the physics of our proposed architecture. In particular, we focus on the left semi-infinite superconductor and on the superconductor at $x=x_{\rm s}$, i.e. only on the left part of Fig.~\ref{fig:setup}. Since there are no ferromagnets under this restricted perspective, right-moving electrons (e) with spin-up can only be Andreev reflected into left-moving holes (h) with spin down and vice versa. We can thus concentrate only on these two particle species and exploit particle-hole symmetry to gain information about the others. While the general results discussed in the following will be obtained by numerically analyzing the scattering problem for all sub-gap energies, here we are primarily interested on the zero-energy Majoranas and we therefore focus on the zero-energy limit. This allows us to show the relation between incoming and outgoing scattering amplitudes (depicted in Fig.~\ref{fig:setup})
	\begin{equation}
	\label{eq:Scat2}
	\left(\begin{matrix}
	b_h \\b_e
	\end{matrix}\right)  = \left(\begin{matrix}
	-i e^{i \chi} \tanh  (\Delta\, l_{\rm S}) & \text{sech} (\Delta\, l_{\rm S})\\
\text{sech} (\Delta\, l_{\rm S}) & 	-i e^{-i \chi} \tanh  (\Delta\, l_{\rm S})
	\end{matrix}\right)\left(\begin{matrix}
	a_e \\ a_h
	\end{matrix}\right),
	\end{equation}
	where $\Delta$, $l_{\rm S}$ and $\chi$ are the proximity induced pairing amplitude, the length and the phase of the finite superconducting region,  respectively. The boundary with the semi-infinite superconductor implies perfect Andreev reflection with $a_e = -i e^{-i\chi_L} b_h$, where $\chi_{\rm L}$ is the phase of the semi-infinite superconductor. The combined effect of the two superconductors leads therefore to perfect Andreev reflection 
	$	b_e = -i e^{-i \chi_{\rm eff}} a_h$ with an effective phase shift
	\begin{equation}
	\label{eq:chi_eff}
	\chi_{\rm eff} (\chi_L,\chi,\Delta\,  l_{\rm S})= \chi - \arg \left[\frac{e^{i \chi}+e^{i \chi_L} \tanh(\Delta \, l_{\rm S})}{e^{i \chi_L}+e^{i \chi} \tanh(\Delta \, l_{\rm S})}\right].
	\end{equation}
	Hence, we conclude that our SSFS system supports zero-energy MBS whenever the condition 
	\begin{equation}
	\label{eq:zero_cond}
	\chi_R = \chi_{\rm eff}(\chi_L,\chi,\Delta\,  l_{\rm S}) -  \pi  
	\end{equation}
	is met. Moreover, the ratio between the scattering amplitudes on both sides of the finite superconductor reads 
	\begin{equation}
	\label{eq:rho}
	\rho= \left|\frac{b_e}{a_e}\right|^2 =  \cosh(2\Delta\,l_{\rm S})+ \sinh(2\Delta\, l_{\rm S}) \cos(\chi_{\rm L}-\chi).
	\end{equation}
	As long as $\Delta\, l_{\rm S}$ is large enough, we can hence trap a zero energy mode either between the two superconductors (when $\chi_{\rm L}-\chi=\pi$) or to the right of the finite one ($\chi_{\rm L}-\chi=0$). The ratio $\rho$ is plotted in Fig.~\ref{fig:fig2}B for $\Delta\, l_{\rm S} = 2$.
	
	Eqs.~\eqref{eq:chi_eff} and \eqref{eq:rho} nicely show the versatility of the SSFS geometry: the phase difference $\chi-\chi_L$ controls the position of one MBS along the edge while the third superconducting phase $\chi_R$ can be used to independently tune the Majorana hybridization. This result is confirmed by the exact numerical study of the SSFS architecture (for all sub-gap energies). In the specific geometry that we consider, the lengths of the middle superconductor and the ferromagnet equal twice the superconducting coherence length $\xi=\Delta_0^{-1}$, i.e. $l_{\rm S}=l_{\rm F}=2\xi$. The total length of the edge is $L=7\xi$ and the middle superconductor and ferromagnet are located at $x_{\rm S}=-x_{\rm F}=-1.5 \xi$, respectively. The hybridization between the two Majoranas is studied in Fig.~\ref{fig:fig2}A, where we plotted $\epsilon_{\rm Maj}$ as a function of the phase differences $\chi-\chi_L$ and $\chi_R-\chi_L$. We considered $m=\Delta=\Delta_0$, while the magnetization angle $\phi$ is not relevant. The red regions indicate a large hybridization, whose actual value exponentially depends on the strength $(m \, l_{\rm F})$ of the ferromagnetic region, as in the standard SFS geometry. By contrast, the white-dashed line highlights points where $\chi_R$ satisfies the zero-energy condition in Eq.~\eqref{eq:zero_cond} and the two Majoranas are thus completely decoupled.

	By changing the phase difference $\chi-\chi_L$ and tuning the third phase $\chi_R$ according to Eq.~\eqref{eq:zero_cond}, it is possible to change the localization of the zero-energy Majoranas, without hybridizing them. In Fig.~\ref{fig:fig2}B we plot the ratio $\rho$ on a logarithmic scale, recalling that $\rho>1$ ($\rho<1$) corresponds to a zero-energy Majorana wavefunction mainly localized to the right (left) of the middle superconductor. In Fig.~\ref{fig:fig2}C we plot the full sub-gap spectrum, which confirms the presence of the zero-energy Majoranas (black dashed line) together with other mid-gap Andreev bound states (ABS) shown in gray. The spectrum is even in the phase difference $\chi-\chi_L$ and symmetric for $\epsilon \to - \epsilon$. We observe that, at $\chi-\chi_L=\pi$, the  energy gap between the zero-energy Majoranas and the first ABS is small and it turns out to be exponentially suppressed in the strength $\Delta\, l_S$ of the middle superconductor \cite{note1}. As a result, there is a trade-off between the possibility to properly localize a Majorana on either sides of the central superconductor [which requires a large $\Delta\, l_S$ according to Eq.~\eqref{eq:rho}] and having a large energy gap between the zero-energy Majoranas and the other mid-gap ABS (which requires a small $\Delta\, l_S$). The numerical results in Fig.~\ref{fig:fig2} shows that $\Delta\, l_S = 2$ is a good compromise.

	In order to better visualize and discuss the localization of the zero-energy MBS, we computed their wavefunctions by using scattering theory [see Appendix \ref{app:B}] and we plot them in Fig.~\ref{fig:fig2}D. In particular, we focus on three parameter configurations which support zero-energy MBS: $\chi-\chi_L = 0$ (star), $\chi-\chi_L = 0.9$ (circle), and $\chi-\chi_L = 1$ (square) while always keeping $\chi_{\rm eff}-\chi_{\rm R}=\pi$. As a function of the phase difference $\chi-\chi_L$, the wavefunction $|\varphi_{0,\alpha}(x)|^2$ of the left Majorana (purple) moves across the finite superconductors, while the other one $|\varphi_{0,\beta}(x)|^2$ (green) is fixed to the right of the ferromagnetic region. Such a behavior directly affects the zero-temperature local density of states (lDOS) of the system [see Appendix~\ref{app:E}]
	\begin{equation}
\mathcal{A}(\omega,x)=\sum_{\epsilon} A_\epsilon(x) \delta(\omega-\epsilon),
\end{equation}
	where the sum is taken over all the bound states energies. Its zero-energy $A_0(x)$ component features indeed two peaks centered over the MBS and which therefore move accordingly, as shown in Fig.~\ref{fig:fig2}D with black dotted lines. We note in passing that our system can be seen as an example of an Andreev molecule \cite{Pillet2018}, where the bound states of each gapless region hybridize with the ones of the neighbor gapless regions.

	\section{Moving Majoranas between different edges} 
	\begin{figure}
		\centering
		\includegraphics[width=1\linewidth]{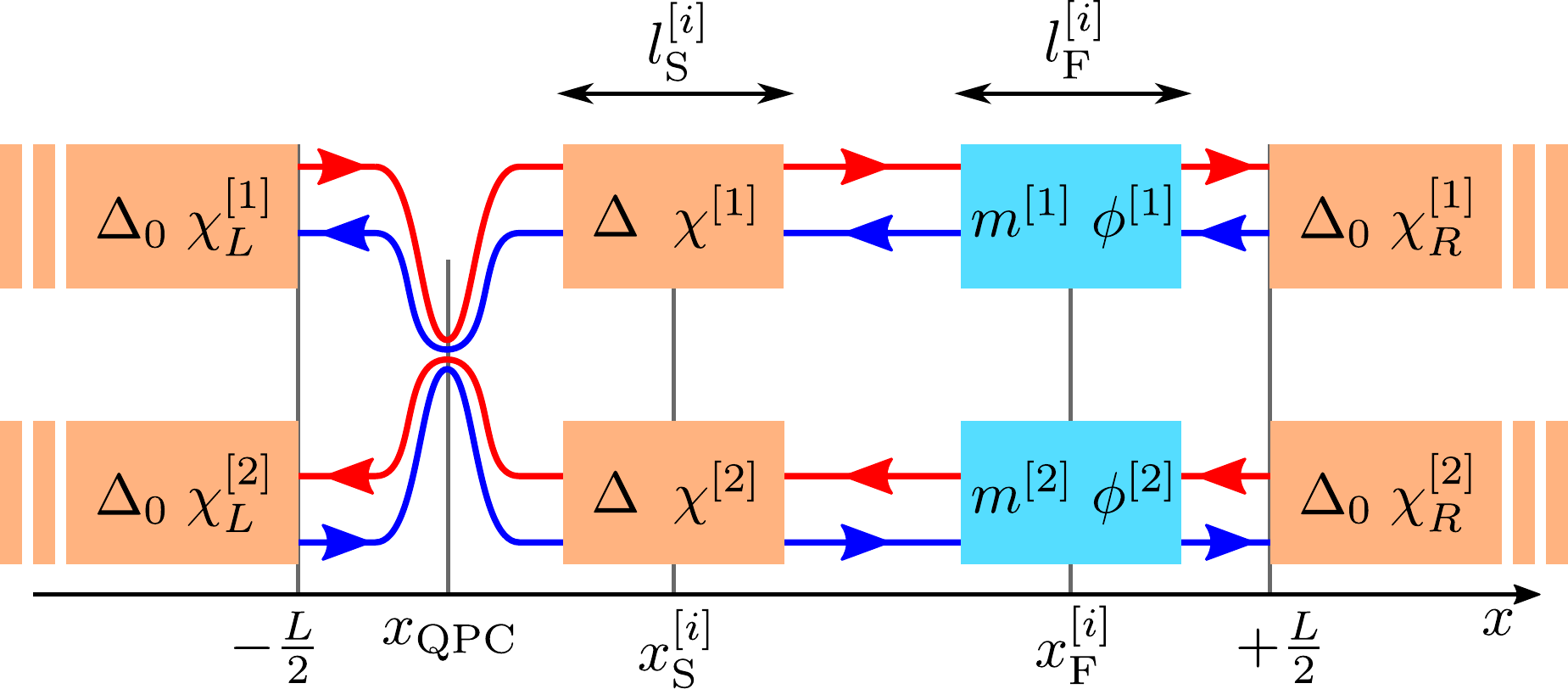}
		\caption{Two helical edges with SSFS geometry, coupled by a QPC at $x=x_{\rm QPC}$. Orange (blue) boxes indicate superconducting (ferromagnetic) gapped regions. Red (blue) lines refers to electron channels with spin up (down). 	
			%(B) Energies $\epsilon_{Exc}$ (red) and $\tilde \epsilon_{Maj}$ (black) as a function of $y$. (C) Zero-energy Majorana wavefunctions $|\varphi_{0,\alpha}(x)|^2$ (green) and $|\varphi_{0,\beta}(x)|^2$ (purple) on the two edges. As $y$ ramps from $0$ to $1$, the purple Majorana moves from the upper edge to the lower one while the green one remains fixed on the upper one. The black dotted lines show (twice) the zero-energy lDOS, i.e. $2 A_0(x)$. The thinner dash-dotted red lines show (four times) the lDOS associated to the first excited state, i.e. $4 A_{\epsilon_{\rm Exc}}(x)$ with $\epsilon_{\rm Exc}>0$. We choose $m^{[1]}=2m^{[2]}$, $\Delta=m^{[2]}=\Delta_0$; the geometry is the same on both edges:  $l_{\rm S}=\xi=\Delta_0^{-1}$, $l_{\rm F}=1.5\xi$, $L=7\xi$, $x_{\rm F}=-x_{\rm S}=1.5 \xi$, $x_{\rm QPC=-3\xi}$. Wavefunctions and lDOS are plotted in units $\xi^{-1}$.%(D) Supercurrent carried by zero-energy MBS which flows throughout the lower right superconductor $(R^{[2]})$, as a function of $y^3$. 
		}
		\label{fig:DE}
	\end{figure}

	The full manipulation of MBS offered by the SSFS architecture represents itself an important achievement. However, a single couple of Majoranas constrained on a one-dimensional edge with open boundaries is not enough to detect and exploit their non-Abelian properties. It is thus necessary to consider systems consisting of several couples of MBS, living on different helical edges which have to be coupled to each other. In this respect, an intriguing and feasible possibility is represented by QPCs, which allow electron tunneling between the edges \cite{Strunz2019}. Remarkably, we prove that QPCs can be used for inter-edge MBS manipulation. Electron tunneling is indeed sensitive to the lDOS, which we just demonstrate to be controllable by moving the MBS along each edge within an SSFS architecture. As a result, the inter-edge coupling provided by the QPC can be effectively and efficiently tuned just by acting on a handful of superconducting phases. We will focus, in particular, on a simple double-edge configuration, which might be realized in current QPC systems \cite{Strunz2019}. 
	
	The system we are considering is shown in Fig.~\ref{fig:DE} and consists of two SSFS edges, coupled by a QPC located between the superconductors at $x=x_{\rm QPC}$. As described in details in Appendix~\ref{app:D}, its Hamiltonian reads \cite{Ferraro2014} ($\vartheta_{R/L}=\pm1$, operators are evaluated at $x= x_{\rm QPC}$)
	\begin{equation}
	\begin{split}
	H_t &= 2 \lambda_{\rm sp} \sum_{\sigma=\uparrow,\downarrow} \psi_{R\sigma}^\dagger\psi_{L\sigma}+ 2 \lambda_{\rm sf} \sum_{r=R,L} \vartheta_r \psi_{r_\uparrow}^\dagger\psi_{r_\downarrow}  + \text{H.c.} .
	\end{split}
	\end{equation}
	It models spin preserving ($\lambda_{\rm sp}$) and spin flipping ($\lambda_{\rm sp}$) tunneling between the channels of the upper edge ($R_\uparrow$ and $L_\downarrow$) and the ones of the lower edge ($R_\downarrow$ and $L_\uparrow$). Starting from the equation of motion, it is possible to derive the $8\times8$ scattering matrix $\mathcal{S}_{\rm QPC}$ associated with the QPC \cite{Ferraro2014} and use the machinery developed in the previous section in order to study the bound states of the whole double-edge system. We mention in passing that the entire setup can also be viewed as a realization of a multi-terminal Josephson junction \cite{Pankratova2018,Riwar2016}. In this section, we will specifically consider the following geometry (see Fig.~\ref{fig:DE} for notation)
	\begin{center}
	\begin{tabular}{lll}
	$ l_{\rm S}^{[1,2]}=2\xi $&$l_{\rm F}^{[1]}=3\xi$ & $l_{\rm F}^{[2]}=2.4\xi$\\[.3em]$ x_{\rm S}^{[1,2]}=-1.5 \xi\quad $&$  x_{\rm F}^{[1]}=1.5 \xi$& $x_{\rm F}^{[2]}=1.8 \xi$\\[.3em]
	$	  L=7\xi$&$ x_{\rm QPC}=-3\xi\;\;$&\\
	\end{tabular}
\end{center}

\begin{figure}
		\centering
		\includegraphics[width=1\linewidth]{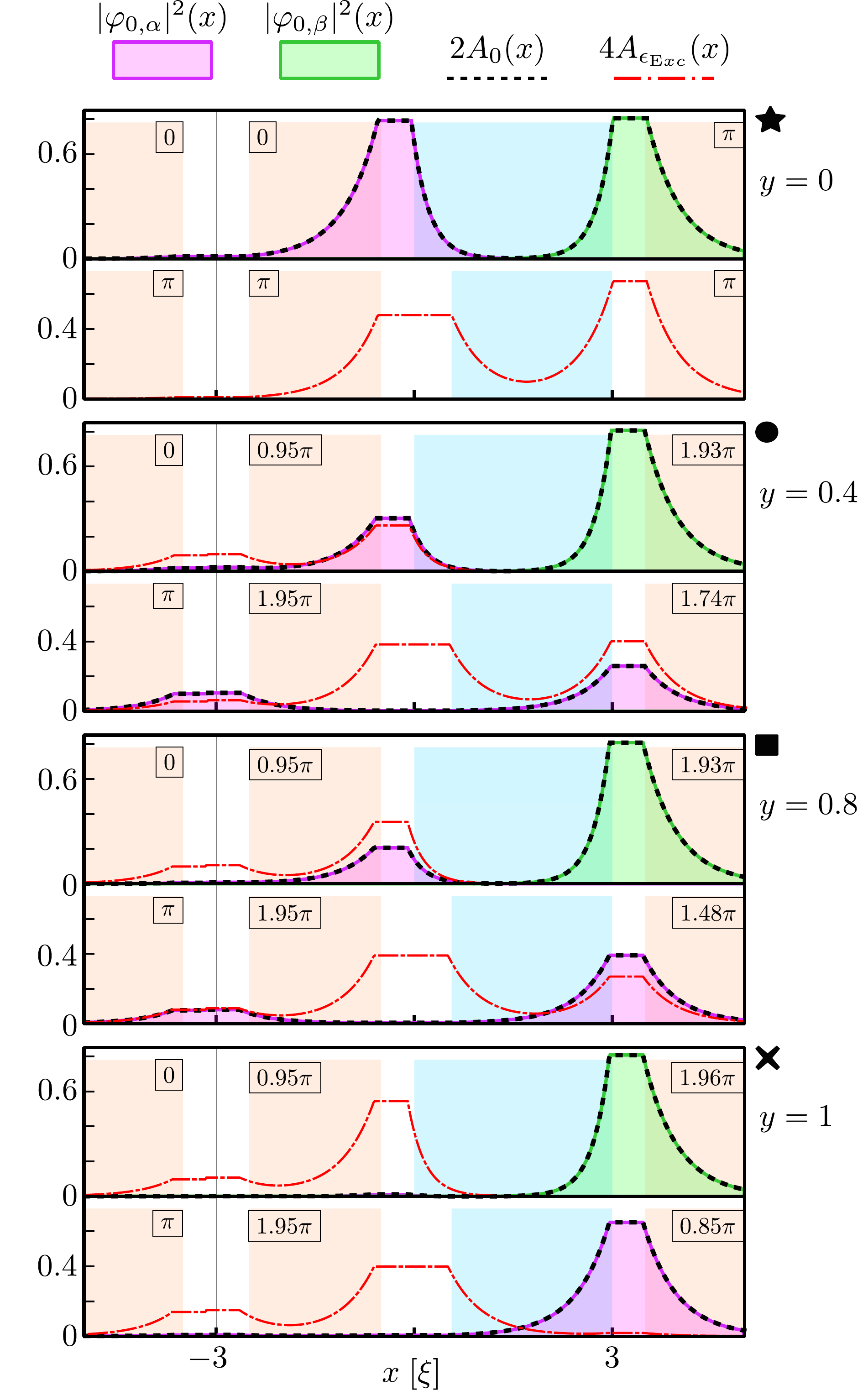}
		\caption{ Zero-energy Majorana wavefunctions $|\varphi_{0,\alpha}(x)|^2$ (green) and $|\varphi_{0,\beta}(x)|^2$ (purple) on the two edges. As $y$ ramps from $0$ to $1$, the purple Majorana moves from the upper edge to the lower one while the green one remains fixed on the upper one. The black dotted lines show (twice) the zero-energy lDOS, i.e. $2 A_0(x)$. The thinner dash-dotted red lines show (four times) the lDOS associated to the first excited state, i.e. $4 A_{\epsilon_{\rm Exc}}(x)$ with $\epsilon_{\rm Exc}>0$. In the little black boxes we report the superconducting phase of each superconductor, assuming $\chi_{\rm L}^{[1]}=0$. We choose $m^{[1]}=2m^{[2]}$, $\Delta=m^{[2]}=\Delta_0$. Wavefunctions and lDOS are plotted in units $\xi^{-1}$.%(D) Supercurrent carried by zero-energy MBS which flows throughout the lower right superconductor $(R^{[2]})$, as a function of $y^3$. 
		}
		\label{fig:DE2b}
	\end{figure}

	We are now able to describe a protocol which allows moving one zero-energy MBS from the upper edge to the lower one. In particular, we start from a configuration which hosts two zero-energy MBS on the upper edge [see Fig.~\ref{fig:DE2b}, top panel (star)] and end with one zero-energy Majorana localized on each edge [see Fig.~\ref{fig:DE2b}, bottom panel (cross)]. Before presenting a quantitative (and numerical) description of the protocol, it is useful to qualitatively show how the initial and final configuration can be achieved by exploiting the QPC and the SSFS architecture. To this end, we observe that when the whole system is tuned such that the lDOS (almost) vanishes in the QPC region, the two edges are effectively decoupled. We can then fully hybridize the MBS on the lower edge ($j=2$), i.e. $\chi_{\rm R}^{[j=2]}\simeq\chi_{\rm eff}(\chi_{\rm L}^{[2]},\chi^{[2]})$, while keeping the Majoranas on the upper edge ($j=1$) at zero-energy, i.e. $\chi_{\rm R}^{[1]}\simeq\pi+\chi_{\rm eff}(\chi_{\rm L}^{[1]},\chi^{[1]})$. This allows us to realize the initial configuration. As for the final one, we tune the system such that the two edges would host two couples of zero-energy MBS localized close to the four semi-infinite superconductors, i.e. $\chi^{[j]}-\chi_{\rm L}^{[j]}\simeq \pi$. In such configuration, the QPC effectively couples and hybridize the two left Majoranas, leaving at zero energy only the ones to the right of the ferromagnets.

		\begin{table}
		\begin{center}
			\begin{tabular}{rcccc}
				edge$\quad$	&$\chi_L^{[j]}$&$\chi^{[j]}$&$\phi^{[j]}$&$\chi_R^{[j]}$\\
				\colrule
				$j=1\quad$& $\;0\;$ & $r(y)$ & $\phi$ & $\pi +\chi_{\rm eff}^{[1]}(y)+f(y)$\\
				$j=2\quad$& $\pi$ &$r(y)+\pi$ & $\quad\phi+\pi\quad$ & $ y\, (\pi +\chi_{\rm eff}^{[2]}(y))+y\, 0.05 \pi+\, \pi$\\
			\end{tabular}
		\end{center}
		\caption{Quantitative description of the protocol which moves one zero-energy Majorana across the QPC.}\label{tab:tab}
	\end{table}

	We quantitatively design the protocol by interpolating between these initial and final configurations. The superconducting phases are tuned, according to Table \ref{tab:tab}, as a function of a single parameter $y$ which ramps from $0$ to $1$. The fixed $\pi$ phase difference between the two edges is essential in order for the QPC to effectively couple the MBS [see Appendix~\ref{app:D}]. The function $r(y)$ reads 
	\begin{equation}
	\label{eq:r}
		r(y)=0.95 \pi  \begin{cases}
		[1-(2y-1)^4]\qquad& 0\leq y< 0.5\\
	1 & 0.5\leq y\leq 1
		\end{cases}
		\end{equation}
	while the continuous function $f(y)$ provides a tiny correction $0\leq f(y) \lesssim0.15 \pi$ to ensure that the two MBS are indeed exactly at zero energy [see Appendix~\ref{app:F}]. The numerical computation of the bound states as a function of $y$ is performed by considering the QPC tunneling amplitudes $\lambda_{\rm sp}=0.1$ and $\lambda_{\rm sf}=0$. The presence of spin-flipping processes does not qualitatively affect the results. We stress that the specific choice of $r(y)$, as well as the presence of a tiny correction $y\, 0.05\pi$ in $\chi_R^{[2]}$, only aim at optimizing some features of the protocol but do not modify the qualitative description given above. Finally, as for the single edge case, the magnetic angle $\phi$ is not relevant. 
	
 Fig.~\ref{fig:DE2b} shows how the Majorana wavefunctions are manipulated during the protocol and represents one of the main results of our work. In particular, we plot the zero-energy MBS wavefunctions $|\varphi_{0,j}(x)|^2$ ($j=\alpha,\beta$) for four different values of $y=0,0.4,0.8,1$ (star, circle, square and cross, respectively). The purple Majorana clearly moves from the upper edge to the lower one, while the green one sticks to the gapless region to the right of the upper ferromagnet. As discussed for the single edge system, the motion of the Majoranas directly affects the lDOS at zero energy $A_0(x)$ (dotted black lines): one of its two peaks indeed moves from one edge to the other one. In the initial configuration $y=0$ (star), when the two edges are almost completely decoupled, it is clear that the system hosts also a couple of midgap bound states at finite energies $\pm \epsilon_{\rm Exc}$, resulting from the full hybridization of the MBS on the lower edge. They give rise to a non-vanishing spectral weight in the lDOS at finite energy $A_{\epsilon _{\rm Exc}}(x)$, which is plotted with dash-dotted red lines. These midgap states evolve as $y$ is ramped from $0$ to $1$ and eventually localize close to the QPC region, as expected from the qualitative description of the final configuration given above. The lDOS can be experimentally probed by performing tunneling spectroscopy \cite{Ren2019}, e.g. by exploiting an additional QPC near pinch-off.
	
\begin{figure}
	\centering
	\includegraphics[width=1\linewidth]{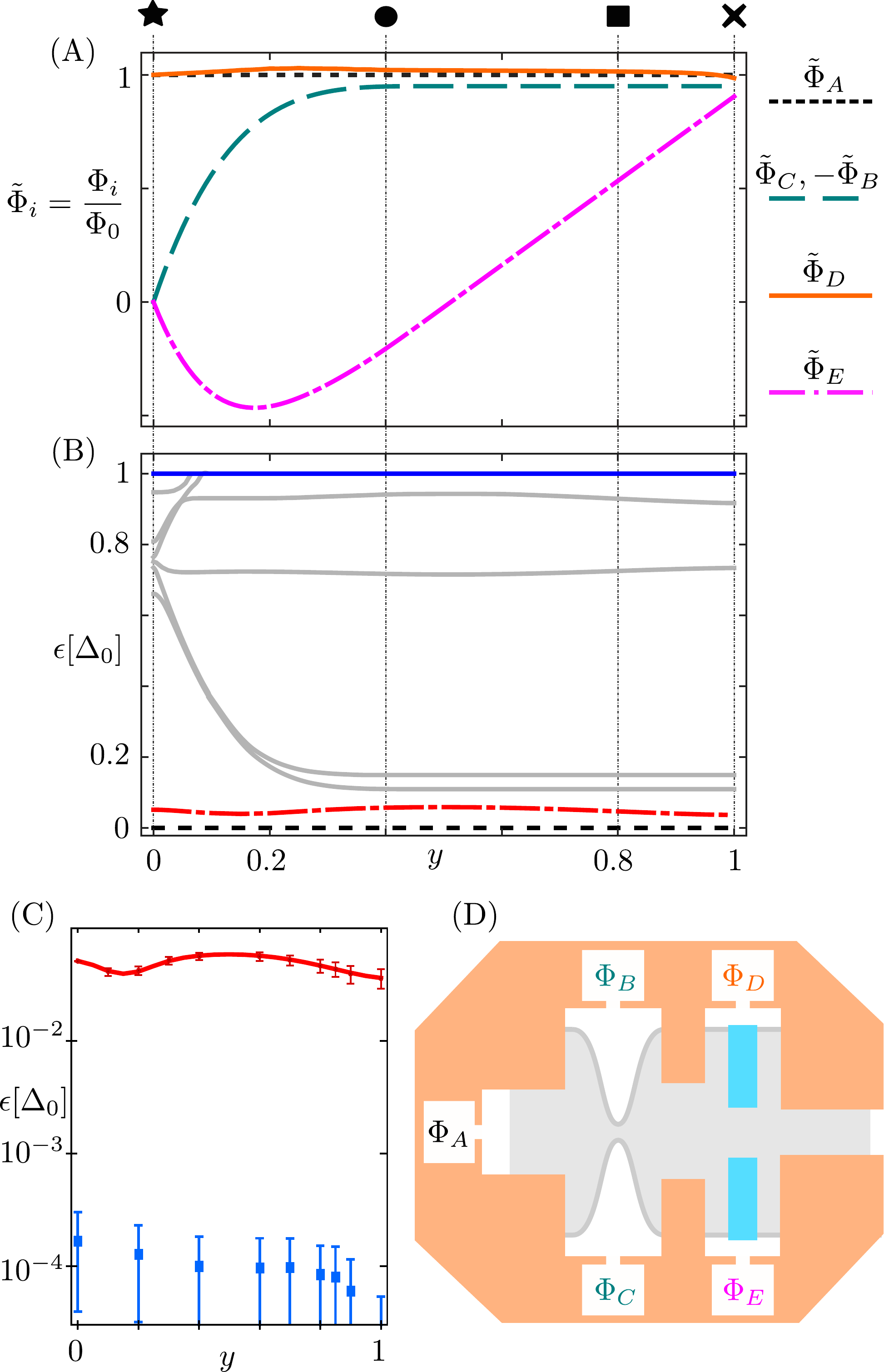}
	\caption{(A) Dependence of the magnetic fluxes (see panel D) on the parameter $y$ throughout the protocol. (B) Sub-gap spectrum of the system as a function of $y$. One can identify the zero-energy Majoranas (black dashed line), the first excited state with energy $\epsilon_{\rm Exc}(y)$ (red dash-dotted line), other mid-gap ABS (gray lines) and the superconducting gap $\Delta_0$ (blue line). The spectrum is symmetric with respect to $\epsilon\to - \epsilon$. The black markers show the specific values of $y$ considered in Fig.~\ref{fig:DE2b}. (C) Logarithmic plot of the low-energy part of the spectrum in presence of finite accuracy in the phase control. Blue and red markers show $\epsilon_{\rm Maj}(y)$ and $\epsilon_{\rm Exc}(y)$, respectively. The error bars represent the standard deviations of every point while the red line shows $\epsilon_{\rm Exc}(y)$ as a reference. 		
		(D) A possible scheme to tune the phase differences across the helical Josephson junction by using magnetic fluxes $\Phi_i$. Different colors refer to superconductors (orange), ferromagnets (light blue) and 2DTI (gray). For all the plots, we choose $m^{[1]}=2m^{[2]}$, $\Delta=m^{[2]}=\Delta_0$.
	}
	\label{fig:DE2}
\end{figure}

	While the two decoupled Majoranas remains at zero-energy throughout all the protocol, it is important to discuss the behavior of other mid-gap states with finite energy. To this end, in Fig.~\ref{fig:DE2}B we plot the sub-gap spectrum as a function of the parameter $y$. Right above the zero-energy Majoranas (black-dashed line), the red dash-dotted line represents the first excited mid-gap state with energy $\epsilon_{\rm Exc}(y)$, whose associated lDOS $A_{\epsilon_{\rm Exc}}(x)$ is plotted in Fig.~\ref{fig:DE2b} (with red dash-dotted line). Importantly, there is always a finite energy gap between this state and the zero-energy Majoranas. In particular, with the protocol described in Table~\ref{tab:tab} and with the interpolating function $r(y)$ in Eq.~\eqref{eq:r}, one has $\epsilon_{\rm Exc}(y)\gtrsim0.04\Delta_0$. Other ABS with higher energy are plotted with gray lines.

	In order to implement the protocol, it is necessary to control the phase differences across the helical Josephson junctions. A possibility is to tune magnetic fluxes \cite{Mi2013,vanHeck2012,Ren2019,Zhou2019} using for example the architecture suggested in Fig.~\ref{fig:DE2}D. In this case, one would have to vary ($\Phi_0 = h/(2e)$ is the flux quantum)
	\begin{align}
 	\Phi_B(y) &=  -\Phi_0\frac{r(y)}{2\pi}
	\\ 
	\Phi_C(y) &= \Phi_0\frac{r(y)}{2\pi}
	\\ 
	\Phi_D(y) &= -\Phi_0\frac{\pi+\chi_{\rm eff}^{[1]}(y)+f(y)-r(y)}{2\pi} 
	\\ 
	\Phi_E(y) &= \Phi_0\frac{y(\pi+\chi_{\rm eff}^{[2]}(y))+y0.05\pi -r(y)}{2\pi}
	\end{align}
	while
	\begin{equation}
	\Phi_A = \frac{\Phi_0}{2}
	\end{equation}
	is fixed. The dependence of these magnetic fluxes on the parameter $y$ is shown in Fig.~\ref{fig:DE2}A. 
	
	It is important to stress that in our setup, the degeneracy of the two zero-energy Majoranas is not topologically protected since its splitting $\epsilon_{\rm Maj}$ is not exponentially small in the deviations of superconducting phases from their optimal values. It is therefore interesting to discuss the robustness of the proposed protocol with respect to a finite accuracy of the phase control. To this end, we add an uncorrelated Gaussian noise with standard deviation $\pi/20$ to each phase independently and then average over many realizations for each given $y$. This finite accuracy affects the energies of all mid-gap states and we focus, in particular, on the zero-energy Majoranas ($\epsilon_{\rm Maj}$) and on the first excited state ($\epsilon_{\rm Exc}(y)$). As shown in Fig.~\ref{fig:DE2}C, a non-optimal control of the superconducting phases leads to a degeneracy splitting $\tilde \epsilon_{\rm Maj} \sim 10^{-4} \, \Delta_0$ (blue markers) and to a modification of $\epsilon_{\rm Exc}(y) \to \tilde \epsilon_{\rm Exc}(y)$ (red markers). The error bars show the standard deviation associated with each point while the red line shows $\epsilon_{\rm Exc}(y)$ as a reference. Importantly, we observe that even with limited accuracy in the phase control (with errors of the order of $\pi/20$), there are still more than two order of magnitude between the two almost-zero-energy Majoranas and the other mid-gap states. As argued in the following paragraphs, this is still compatible with a successful implementation of the protocol.

		\section{Discussion and conclusions}
		
		Our proposal, although challenging, is experimentally feasible. Indeed, it has been shown that it is possible to proximitize the helical edges by using Al superconducting contacts \cite{Bocquillon2016, Wiedenmann2016, Bocquillon2018}. The superconducting coherence length is $\xi \sim 0.6\div2\, \mu m$ while the induced superconducting gap is $\Delta \sim 90 \, \mu eV$. Moreover, the ballistic mean free path on the helical edges is estimated to be $l_{\textrm b} \sim 2.4 \mu m$. Regarding quantum point contacts, the physical size of the first-realized QPC in HgTe 2DTIs is around $l_{\textrm QPC}\sim 0.5\, \mu m$ \cite{Strunz2019}. It is therefore reasonable to assume our helical gapless regions, whose lengths are of the order of $\xi$, to be wide enough to accommodate a QPC but short enough to assume ballistic transport. As for the ferromagnetic regions, we assumed $m \sim \Delta_0$ which corresponds to $m/\mu_0\sim 1T$, where $\mu_0$ is the Bohr magneton.
		
		The energy gap between the Majoranas and the first excited mid-gap states is of the order of $\epsilon_{\rm Exc}\sim 4\, \mu V$, which corresponds to a temperature scale $T_{\rm Exc} =\epsilon_{\rm Exc}/k_{\rm B}\sim 5\,\cdot 10^{-2}\; K$ and to a time scale $t_{\rm Exc} = h/\epsilon_{\rm Exc}\sim 1\, ns$. As long as the protocol is performed over a time span $t_{p}$ much greater than $t_{\rm Exc}$, we can safely assume an adiabatic evolution of the system. In general, upper bounds to the total moving time $t_p$ exist as well. One might be represented by the quasiparticle poisoning time $t_{\textrm qp}$, which strongly depends on the superconducting system \cite{Sanchez-Barriga2012,Budich2012}, but should in any case exceed hundreds of nanoseconds \cite{Rainis2012}. Another time scale $t_{\rm GS}$ arises when the groundstate degeneracy is split, for example, because of a finite accuracy in tuning the superconducting phases. In this case, for an accuracy up to  $\pi/20$, one has $t_{\rm GS}=h/\tilde \epsilon_{\rm Maj}\sim 4\, 10^2 \, ns$. If $t_p$ exceeds this time scale, the difference in the dynamical phases gained by the two lowest energy states is not negligible anymore. In general, however, it is possible to operate at an intermediate regime $t_{\rm Exc}\ll t_p \ll t_{\rm GS},t_{\rm qp}$.

	The proposed SSFS geometry allows us to achieve a promising tunability of a single couple of MBS on the helical edge of a 2DTI. In particular, we can independently move the Majoranas along the edge and tune their hybridization. While already interesting on its own, this architecture develops its full potential when implemented on multiple edges, coupled via a QPC. The interplay between inter-edge electron tunneling and intra-edge Majorana manipulation results in the possibility to fully control the position of single MBS over the whole system, just by tuning a handful of superconducting phases. In particular, the present proposal deals with a simple double-edge configuration, which might be realized in current QPC systems on 2DTI. We invent a protocol which allows us to move one MBS from the upper to the lower edge, while leaving the other MBS untouched. This change in the position of a MBS, controlled by tuning four superconducting phase differences, directly affects the zero-energy lDOS of the system.
	
	The combination of the SSFS geometry with the QPC dramatically enhances the capabilities of 2DTI systems in manipulating MBS. Despite the lack of topological protection, our proposal paves the way for a challenging but feasible implementation of braiding schemes which can reveal the non-Abelian nature of Majoranas. 
	
	\begin{acknowledgments}
		We would like to thank  M. Carrega, C. Fleckenstein, F. Keidel, M. Stehno, and S. Zhang for fruitful discussion. We acknowledge support by the W\"urzburg-Dresden Cluster of Excellence on Complexity and Topology in Quantum Matter (EXC 2147, project-id 39085490).
	\end{acknowledgments}
	
	\appendix
	
	\section{BdG Hamiltonian}
	\label{app:A}
	\noindent
	We first analyze the Hamiltonian of a single helical edge, in presence of superconducting and ferromagnetic regions. The latter can be conveniently expressed as
	\begin{equation}
	H = \frac{1}{2} \int dx\; \Psi^\dagger \; \mathcal{H}_{\rm BdG} \Psi,
	\end{equation}
	with the Nambu spinor
	\begin{equation}
	\label{eq:Nambu}
	\Psi = (\psi_{\uparrow},\psi_{\downarrow},\psi_{\downarrow}^\dagger,-\psi_{\uparrow}^\dagger)^T
	\end{equation}
	and the Bogoliubov-de Gennes Hamiltonian
	\begin{equation}
	\label{eq:app:BdG}
	\mathcal{H}_{\rm BdG} (x) = \pm p_x  \tau_3 \sigma_3+ \vec{m}(x) \cdot \vec{\sigma} + \vec{\Delta}(x) \cdot \vec{\tau}-\mu(x) \tau_3
	\end{equation} 
	The sign in front of the momentum operator select the helicity of the edge: a plus (minus) sign corresponds to right-moving electrons with spin up (down) and vice-versa. The Pauli matrices $\vec{\sigma}=(\sigma_1,\sigma_2,\sigma_3)$ [ $\vec{\tau}=(\tau_1,\tau_2,\tau_3)$] act on spin (particle-hole) space, $p_x = -i \partial_x$ is the momentum operator, $\mu(x)$ the chemical potential and we have set both $\hbar = 1$ and the Fermi velocity $v_{\rm F} =1$.  The superconducting and ferromagnetic regions are described by non-vanishing
	\begin{align}
	\vec{\Delta} &= (\Delta \cos\chi, \Delta \sin\chi, 0)^T\\
	\vec{m}&=(m \cos\phi, m\sin\phi, m_z)^T,
	\end{align}
	respectively. The Hamiltonian can be diagonalized as%(here we assume for simplicity that no degeneracies are present, we will deal with them later)
	\begin{equation}
	\label{eq:BdGdiag}
	H=\sum_{\epsilon\geq 0} \sum_j \epsilon \gamma_{\epsilon,j}^\dagger  \gamma_{\epsilon,j},
	\end{equation}
	where the operators $\gamma_{\epsilon_n,j}^\dagger $ and $\gamma_{\epsilon,j}$ create and annihilate a fermionic quasiparticle with energy $\epsilon$, respectively \cite{Crepin2014}. The index $j$ takes into account possible degeneracies. The energy of the groundstate(s) has been set to zero. The diagonalized form \eqref{eq:BdGdiag} is achieved by exploiting the ansatz on the Nambu spinor
	\begin{equation}
	\label{eq:ansatz}
	\Psi(x) = \sum_{\epsilon\geq 0}  \sum_j \varphi_{\epsilon,j}(x) \gamma_{\epsilon,j} + \left[\mathcal{C}\varphi_{\epsilon,j}\right](x) \gamma^\dagger_{\epsilon,j},
	\end{equation} 
	where
	\begin{equation}
	\varphi_{\epsilon,j} = \left(u_{\epsilon,j,\uparrow},u_{\epsilon,j,\downarrow},v_{\epsilon,j,\downarrow},v_{\epsilon,j,\uparrow}\right)^T
	\end{equation}
	is a solution of the BdG equation
	\begin{equation}
	\mathcal{H}_{\rm BdG} \; 	\varphi_{\epsilon,j}  = \epsilon 	\varphi_{\epsilon,j} 
	\end{equation}
	and $\mathcal{C} = \mathcal{K}  \tau_2\sigma_2$, with $\mathcal{K} $ the complex conjugation, is the charge-conjugation operator \cite{Crepin2014}. The BdG Hamiltonian \eqref{eq:app:BdG} features a built-in particle hole symmetry
	\begin{equation}
	\mathcal{C} \mathcal{H}_{\rm BdG} \mathcal{C}^{-1} =- \mathcal{H}_{\rm BdG}
	\end{equation} 
	which implies
	\begin{equation}
	\mathcal{H}_{\rm BdG}\varphi_{\epsilon,j} = \epsilon \varphi_{\epsilon,j} \quad \Rightarrow \quad \mathcal{H}_{\rm BdG}(\mathcal{C} \varphi_{\epsilon,j}) =  -\epsilon (\mathcal{C} \varphi_{\epsilon,j}).
	\end{equation}
	When degeneracies are not present, one can suppress the index $j$ and exploit the identity (up to a global phase)
	\begin{equation}
	\label{eq:uv_epsilon}
	\left(
	\begin{matrix}
	u_\uparrow(\epsilon)\\
	u_\downarrow(\epsilon)\\
	v_\downarrow(\epsilon)\\
	v_\uparrow(\epsilon)
	\end{matrix}\right)=\left(
	\begin{matrix}
	-v_\uparrow(-\epsilon)^*\\
	v_\downarrow(-\epsilon)^*\\
	u_\downarrow(-\epsilon)^*\\
	-u_\uparrow(-\epsilon)^*\\
	\end{matrix}\right).
	\end{equation}
	
	\subsection{Bogoliubov equations}
	Solving the Bogoliubov-de Gennes equation, this allows us to determine the expressions for the wavefunction components $u_{\uparrow,\downarrow}(\epsilon,x)$ and $v_{\uparrow,\downarrow}(\epsilon,x)$ of the bound states. To this end, we transform the equation
	\begin{widetext}
	\begin{equation}
	\label{eq:BdGeq}
	\pm (-i \partial_x)\left(
	\begin{matrix}
	1&0&0&0\\
	0&-1&0&0\\
	0&0&-1&0\\
	0&0&0&1
	\end{matrix}\right)
	\left(
	\begin{matrix}
	u_\uparrow\\
	u_\downarrow\\
	v_\downarrow\\
	v_\uparrow
	\end{matrix}\right)+
	\left(
	\begin{matrix}
	-\mu-\epsilon+m_z&m\, e^{-i\phi}&\Delta\, e^{-i\chi}&0\\
	m\, e^{i\phi}&-\mu-\epsilon-m_z&0&\Delta\, e^{-i\chi}\\
	\Delta\, e^{i\chi}&0&\mu-\epsilon+m_z&m\, e^{-i\phi}\\
	0&\Delta\, e^{i\chi}&m\, e^{i\phi}&\mu-\epsilon-m_z
	\end{matrix}\right)
	\left(
	\begin{matrix}
	u_\uparrow\\
	u_\downarrow\\
	v_\downarrow\\
	v_\uparrow
	\end{matrix}\right) = 0
	\end{equation}
	distinguishing between the two helicities $\pm$, as
	\begin{align}
	\text{Helicity } +: \qquad &  (-i \partial_x)
	\left(
	\begin{matrix}
	u_\uparrow\\
	-u_\downarrow\\
	-v_\downarrow\\
	v_\uparrow
	\end{matrix}\right)+
	\left(
	\begin{matrix}
	-\mu-\epsilon+m_z&-m\, e^{-i\phi}&-\Delta\, e^{-i\chi}&0\\
	m\, e^{i\phi}&\mu+\epsilon+m_z&0&\Delta\, e^{-i\chi}\\
	\Delta\, e^{i\chi}&0&-\mu+\epsilon-m_z&m\, e^{-i\phi}\\
	0&-\Delta\, e^{i\chi}&-m\, e^{i\phi}&\mu-\epsilon-m_z
	\end{matrix}\right)
	\left(
	\begin{matrix}
	u_\uparrow\\
	-u_\downarrow\\
	-v_\downarrow\\
	v_\uparrow
	\end{matrix}\right) = 0 \\
	\text{Helicity } -: \qquad & 
	(-i \partial_x)
	\left(
	\begin{matrix}
	u_\downarrow\\
	-u_\uparrow\\
	-v_\uparrow\\
	v_\downarrow
	\end{matrix}\right)+
	\left(
	\begin{matrix}
	-\mu-\epsilon-m_z&-m\, e^{+i\phi}&-\Delta\, e^{-i\chi}&0\\
	m\, e^{-i\phi}&\mu+\epsilon-m_z&0&\Delta\, e^{-i\chi}\\
	\Delta\, e^{i\chi}&0&-\mu+\epsilon+m_z&m\, e^{i\phi}\\
	0&-\Delta\, e^{i\chi}&-m\, e^{-i\phi}&\mu-\epsilon+m_z
	\end{matrix}\right)
	\left(
	\begin{matrix}
	u_\downarrow\\
	-u_\uparrow\\
	-v_\uparrow\\
	v_\downarrow
	\end{matrix}\right) = 0.
	\end{align}
	\end{widetext}
	Hence, we further analyze only one helicity, say the $+$ one, since the solutions for the other one can be easily obtained by implementing the transformation
	\begin{align}
	m_z &\leftrightarrow -m_z\\
	\phi &\leftrightarrow -\phi\\
	\uparrow &\leftrightarrow \;  \downarrow
	\end{align}
	
	\section{Scattering matrices}
	\label{app:B}
	In order to identify the presence of mid-gap bound states and to study their wavefunction, we employ scattering theory. To this end, we have to associate a scattering matrix to each ferromagnetic and superconducting region. 
	
	\begin{figure}
		\centering
		\includegraphics[width=0.9\linewidth]{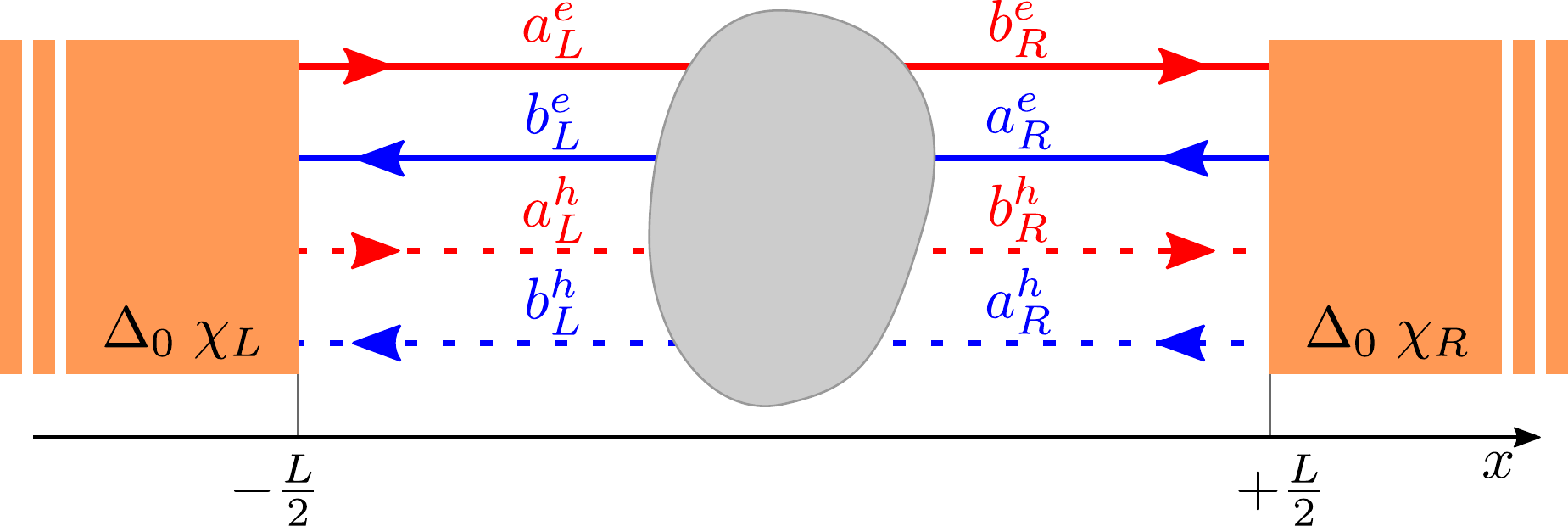}
		\caption{A generic scattering region (gray) in between two semi-infinite superconductors (orange). Solid (dashed) lines refer to electron (hole) channels. Different colors refer to opposite spin polarizations. The scattering amplitudes are labeled in the following way: $a$ is the amplitude associated with electrons ($e$) or holes ($h$) incoming from the left ($L$) or from the right ($R$) of the gray area. Analogously, $b$ is the amplitude associated with electrons ($e$) or holes ($h$) outgoing to the left ($L$) or to the right ($R$).}
		\label{fig:smfig1}
	\end{figure}

	\subsection{Ferromagnetic gapped region}
	Let us first consider a ferromagnetic gapped region, centered in $x_0$ with width $l$. We set $\Delta=0$ and focus only on states within the gap, i.e. with energies $|m|>|\epsilon+\mu|$. The BdG equation \eqref{eq:BdGeq} is then block diagonal and we can focus at first on the electronic sector
	\begin{equation}
	(-i \partial_x)
	\left(
	\begin{matrix}
	u_\uparrow\\
	-u_\downarrow
	\end{matrix}\right)+
	\left(
	\begin{matrix}
	-\mu-\epsilon+m_z&-m\, e^{-i\phi}\\
	m\, e^{i\phi}&\mu+\epsilon+m_z\\
	\end{matrix}\right)
	\left(
	\begin{matrix}
	u_\uparrow\\
	-u_\downarrow
	\end{matrix}\right) = 0.
	\end{equation}
	It admits exponential solutions at a given energy $\epsilon$ which read
	\begin{align}
	u_\uparrow &=   e^{- i m_z x-i\phi} \left[(-i\kappa-(\epsilon+\mu)) A_1 e^{- x\kappa}  +  (i\kappa-(\epsilon+\mu)) A_2 e^{ x\kappa}\right],\\
	u_\downarrow&= - m e^{- i m_z x } \left[  A_1 e^{- x\kappa} +A_2 e^{ x\kappa} \right],
	\end{align}
	with generic complex parameters $A_i$ and $\kappa =\sqrt{m^2-(\epsilon+\mu)^2} $. By contrast, in a gapless region with no in-plane magnetization $m=0$, the plane wave solutions read
	\begin{align}
	u_\uparrow &=   C_1 \; e^{i(\tilde \mu+\epsilon-\tilde m_z)x},\\
	u_\downarrow&= C_2 \; e^{-i(\tilde\mu+\epsilon+\tilde m_z)x}.
	\end{align}
	For the sake of generality, we consider non-vanishing chemical potential ($\tilde \mu$) and magnetization along $z$ ($\tilde m_z$), even outside the ferromagnetic region. By matching these solutions on the boundary of the ferromagnetic region, we can construct the scattering matrix $\mathcal{S}$ which describes it. The latter relates the incoming scattering amplitudes $\vec{a}$ with the outgoing ones $\vec{b}$ %\footnote{To be more precise, } 
	(see Fig.~\ref{fig:smfig1})
	\begin{equation}
	\left(\begin{matrix}
	b^e_L\\[.25em]b^e_R\\[.25em]b^h_L\\[.25em]b^h_R
	\end{matrix}
	\right) = \mathcal{S} \left(\begin{matrix}
	a^e_L\\[.25em]a^e_R\\[.25em]a^h_L\\[.25em]a^h_R
	\end{matrix}
	\right).
	\end{equation}
	In particular, the continuity constraint for an electronic scattering from the left reads
	\begin{widetext}
	\begin{equation}
	\label{eq:hel1_L}
	\begin{cases}
	u_\uparrow (x_0-l/2) = e^{+ i(\tilde \mu+\epsilon-\tilde m_z)(x_0-l/2)} &=  e^{- i m_z (x_0-l/2)-i\phi} \left[(-i\kappa-(\epsilon+\mu)) A_{1_L} e^{- (x_0-l/2)\kappa}  +  (i\kappa-(\epsilon+\mu)) A_{2_L} e^{+ (x_0-l/2)\kappa}\right]\\
	u_\downarrow (x_0-l/2) = r_L^e e^{- i(\tilde \mu+\epsilon+\tilde m_z)(x_0-l/2)} &=  - m e^{- i m_z (x_0-l/2) } \left[  A_{1_L} e^{-(x_0-l/2)\kappa} +A_{2_L} e^{+ (x_0-l/2)\kappa} \right]\\
	u_\uparrow (x_0+l/2) = t_L^e e^{+ i(\tilde \mu+\epsilon-\tilde m_z)(x_0+l/2)} &=  e^{- i m_z (x_0+l/2)-i\phi} \left[(-i\kappa-(\epsilon+\mu)) A_{1_L} e^{-(x_0+l/2)\kappa}  +  (i\kappa-(\epsilon+\mu)) A_{2_L} e^{+ (x_0+l/2)\kappa}\right]\\
	u_\downarrow (x_0+l/2) = 0 &= - m e^{- i m_z (x_0+l/2) } \left[  A_{1_L} e^{- (x_0+l/2)\kappa} +A_{2_L} e^{+ (x_0+l/2)\kappa} \right].\\
	\end{cases}
	\end{equation}
	An electrinic scattering from the right is instead described by
	\begin{equation}
	\label{eq:hel1_R}
	\begin{cases}
	u_\uparrow (x_0-l/2) = 0 &=  e^{- i m_z (x_0-l/2)-i\phi} \left[(-i\kappa-(\epsilon+\mu)) A_{1_R} e^{- (x_0-l/2)\kappa}  +  (i\kappa-(\epsilon+\mu)) A_{2_R} e^{+ (x_0-l/2)\kappa}\right]\\
	u_\downarrow (x_0-l/2) = t_R^e e^{- i(\tilde \mu+\epsilon+\tilde m_z)(x_0-l/2)} &=  - m e^{- i m_z (x_0-l/2) } \left[  A_{1_R} e^{-(x_0-l/2)\kappa} +A_{2_R} e^{+ (x_0-l/2)\kappa} \right]\\
	u_\uparrow (x_0+l/2) = r_R^e e^{+ i(\tilde \mu+\epsilon-\tilde m_z)(x_0+l/2)} &=  e^{- i m_z (x_0+l/2)-i\phi} \left[(-i\kappa-(\epsilon+\mu)) A_{1_R} e^{-(x_0+l/2)\kappa}  +  (i\kappa-(\epsilon+\mu)) A_{2_R} e^{+ (x_0+l/2)\kappa}\right]\\
	u_\downarrow (x_0+l/2) =  e^{- i(\tilde \mu+\epsilon+\tilde m_z)(x_0+l/2)} &= -m e^{- i m_z (x_0+l/2) } \left[  A_{1_R} e^{- (x_0+l/2)\kappa} +A_{2_R} e^{+ (x_0+l/2)\kappa} \right]\\
	\end{cases}
	\end{equation}
	By solving these systems, we obtain 
	\begin{align}
	r_L^e &= m e^{-i(l-2x_0)(\epsilon+\tilde \mu)}e^{+i\phi} \left(e^{2l\kappa}-1\right)\Xi^{-1}\\
	t_L^e &= 2i\kappa e^{-il(\epsilon+\tilde \mu)}e^{-i l (m_z-\tilde m_z)} \; e^{l\kappa}\,\Xi^{-1}\\
	r_R^e &= m e^{-i(l+2x_0)(\epsilon+\tilde \mu)}e^{-i\phi} \left(e^{2l\kappa}-1\right)\Xi^{-1}\\
	t_R^e &= 2i\kappa e^{-il(\epsilon+\tilde \mu)}e^{+i l (m_z-\tilde m_z)} \; e^{l\kappa}\,\Xi^{-1}
	\end{align}
	with 
	\begin{equation}
	\Xi = (\epsilon+\mu)(e^{2l\kappa}-1)+i\kappa(1+e^{2l\kappa}).
	\end{equation}
	Inside the gapped region, the wavefunction is described by
	\begin{align}
	A_{1_L} &= -\exp\left[i\phi + 2\kappa l + (x_0-l/2)(\kappa + i (\epsilon+\tilde \mu + m_z-\tilde m_z))\right]\;  \Xi^{-1}\\
	A_{2_L} &= -A_{1_L} \; \exp\left[-\kappa (2x_0+l)\right]\\
	A_{1_R} &= -	A_{2_R}\; \exp\left[-\kappa (2x_0-l)\right] (\epsilon+u+i\kappa)(\epsilon+u-i\kappa)^{-1} \\
	A_{2_R} &= -(\epsilon+\mu+i\kappa)m^{-1}\exp\left[ -i (x_0+l/2) (\epsilon+\tilde \mu - m_z+\tilde m_z))-x_0\kappa+3l\kappa/2\right]\;  \Xi^{-1} 
	\end{align}
	\end{widetext}
	These results allow us to construct the full scattering matrix by exploiting the particle-hole symmetry of the Hamiltonian (in particular Eq.~\eqref{eq:uv_epsilon})
	\begin{equation}
	\mathcal{S}_{\rm F} (\epsilon)= 
	\left(
	\begin{matrix}
	r_L^e(\epsilon) & t_R^e(\epsilon)&0&0\\
	t_L^e(\epsilon) & r_R^e(\epsilon)&0&0\\
	0&0&-r_L^e(-\epsilon)^* & t_R^e(-\epsilon)^*\\
	0&0&t_L^e(-\epsilon)^* & -r_R^e(-\epsilon)^*
	\end{matrix}\right)
	\end{equation}
	\subsection{Superconducting gapped region}
	Here we consider a superconducting region, with $m=m_z=0$, centered in $x_0$ and with width $l$. The BdG can be cast in a block diagonal form and we can thus focus again only on two variables
	\begin{equation}
	(-i \partial_x)
	\left(
	\begin{matrix}
	u_\uparrow\\
	-v_\downarrow
	\end{matrix}\right)+
	\left(
	\begin{matrix}
	-\mu-\epsilon+m_z&-\Delta\, e^{-i\chi}\\
	\Delta\, e^{i\chi}&-\mu+\epsilon-m_z\\
	\end{matrix}\right)
	\left(
	\begin{matrix}
	u_\uparrow\\
	-v_\downarrow
	\end{matrix}\right) = 0.
	\end{equation}
	Such an equation is equivalent to the one for the ferromagnet provided that the following identifications are made
	\begin{center}\begin{tabular}{ccc}
			\toprule
			\hspace{.25cm} 	Ferromagnet \hspace{.25cm} && \hspace{.25cm}Superconductor\hspace{.25cm} \\
			\colrule
			
			$m$ & $\leftrightarrow$ &$\Delta$\\
			$\phi$ &$\leftrightarrow$ & $\chi$\\
			$m_z$ & $\leftrightarrow$ &$-\mu$\\
			$\mu$ &$\leftrightarrow$ & $0$\\
			$\tilde m_z$ &$\leftrightarrow$ & $-\tilde \mu$\\
			$\tilde \mu$ & $\leftrightarrow$ &$-\tilde m_z$ \\
			$u_\downarrow$&$\leftrightarrow$ &$v_\downarrow$\\
			\botrule
		\end{tabular}
	\end{center}
	We can thus immediately obtain the results
	\begin{align}
	\label{eq:rL1}
	r_L^{(1)} &= \Delta e^{-i(l-2x_0)(\epsilon-\tilde m_z)}e^{+i\chi} \left(e^{2l\nu}-1\right)\Theta^{-1}\\
	t_L^{(1)} &= 2i\nu e^{-il(\epsilon-\tilde m_z)}e^{i l( \mu-\tilde \mu)} \; e^{l\nu}\,\Theta^{-1}\\
	r_R^{(1)} &= \Delta e^{-i(l+2x_0)(\epsilon-\tilde m_z)}e^{-i\chi} \left(e^{2l\nu}-1\right)\Theta^{-1}\\
	\label{eq:tR1}
	t_R^{(1)} &= 2i\nu e^{-il(\epsilon-\tilde m_z)}e^{-i l (\mu-\tilde \mu)} \; e^{l\nu}\,\Theta^{-1}
	\end{align}
	with 
	\begin{equation}
	\Theta = \epsilon(e^{2l\nu}-1)+i\nu(1+e^{2l\nu}).
	\end{equation}
	and $\nu=\sqrt{\Delta^2-\epsilon^2}$. Again, we are only considering states within the gap $|\epsilon|< |\Delta|$. The tilde quantities $\tilde \mu$ and $\tilde m_z$ refer to chemical potential and magnetization along $z$ in the gapless regions. By exploiting particle-hole symmetry, %\footnote{One can show that the scattering matrix obeys
	%	S(\epsilon) = - \left[(\tau_x\sigma_z) S(-\epsilon)^* %(\tau_x\sigma_z)\right]
	%\end{equation}}
	we get the scattering matrix
	\begin{equation}
	\mathcal{S}_{\rm S} (\epsilon)= 
	\left(
	\begin{matrix}
	0&  t_R^{(1)}(-\epsilon)^*& -r_L^{(1)}(-\epsilon)^*&0\\
	t_L^{(1)}(\epsilon) & 0&0&r_R^{(1)}(\epsilon)\\
	r_L^{(1)}(\epsilon)&0&0 &  t_R^{(1)}(\epsilon)\\
	0& - r_R^{(1)}(-\epsilon)^*& t_L^{(1)}(-\epsilon)^*&0
	\end{matrix}\right).
	\end{equation}
	
	\subsection{Multiple scattering regions: transfer matrices}
	In order to deal with multiple scattering barriers, it is necessary to work with transfer matrices $\mathcal{T}$, which relate the left scattering amplitudes with the right ones. For example, referring to the situation depicted in Fig.~\ref{fig:smfig1}, we would have
	\begin{equation}
	\left(\begin{matrix}
	b^e_R\\[.25em]a^e_R\\[.25em]b^h_R\\[.25em]a^h_R
	\end{matrix}
	\right) = \mathcal{T} \left(\begin{matrix}
	a^e_L\\[.25em]b^e_L\\[.25em]a^h_L\\[.25em]b^h_L
	\end{matrix}
	\right).
	\end{equation}
	In presence of subsequent scattering regions $1$ and $2$, the combined transfer matrix is just the product of the individual transfer matrices $\mathcal{T}_{\rm res} =\mathcal{T}_{\rm 2}\mathcal{T}_{\rm 1}$. Transfer matrices are univocally related to scattering matrices. 
	
	\subsection{Semi-infinite superconductors}
	We next discuss the properties of semi-infinite superconductors, i.e. with a width $l\to\infty$, as sketched in Fig.~\ref{fig:smfig1}. Let us focus at first on the left one, with pairing potential $\Delta_0$, phase $\chi_L$, and spatial extension $-\infty<x<-L/2$. According to Eqs. (\ref{eq:rL1} - \ref{eq:tR1}), particles impinging from the right are completely Andreev reflected back since $t_R^{(1)}\to 0$ and $|r_R^{(1)}|\to1$. In particular, for the $+$ helicity, one has
	\begin{align}
	u_\uparrow &= r_R^{(1)}(\epsilon)\;  v_\downarrow = e^{iL(\epsilon-\tilde m_z)}\, e^{-i\chi_L} \, \frac{\Delta_0}{\epsilon+i\sqrt{\Delta_0^2-\epsilon^2}} v_\downarrow\notag =\\&= e^{iL\epsilon}\, \exp\left[-i\chi_L-i L \tilde m_z -i\arccos\left(\tfrac{\epsilon}{\Delta_0}  \right) \right] v_\downarrow\\
	v_\uparrow &= -r_R^{(1)}(-\epsilon)^*\;  u_\downarrow\notag =\\& = 
	- e^{iL\epsilon}\, \exp\left[+i\chi_L+i L \tilde m_z +i\arccos\left(\tfrac{-\epsilon}{\Delta_0}  \right) \right] u_\downarrow 
	\end{align}
	The same reasoning applies to the semi-infinite superconductor which extends from $L/2<x<\infty$ with superconducting phase $\chi_R$. Then, the perfect Andreev reflection takes the form 
	\begin{align}
	v_\downarrow &= r_L^{(1)}(\epsilon)\;  u_\uparrow \notag =\\&= e^{iL\epsilon}\, \exp\left[+i\chi_R-i L \tilde m_z -i\arccos\left(\tfrac{\epsilon}{\Delta_0}  \right) \right] u_\uparrow\\
	u_\downarrow &= -r_L^{(1)}(-\epsilon)^*\;  v_\uparrow \notag =\\&=  -e^{iL\epsilon}\, \exp\left[-i\chi_R+i L \tilde m_z +i\arccos\left(\tfrac{-\epsilon}{\Delta_0}  \right) \right]  v_\uparrow.
	\end{align}
	Results for the opposite helicity are simply obtained by exchanging $\uparrow\;  \leftrightarrow\; \downarrow$ and by changing the signs of $\tilde m_z$.
	
	If the generic scattering region, depicted in gray in Fig.~\ref{fig:smfig1}, is described by the scattering matrix $\mathcal{S}_{\rm in}$, which relates
	\begin{equation}
	\left(\begin{matrix}
	b^e_L\\[.25em]b^e_R\\[.25em]b^h_L\\[.25em]b^h_R
	\end{matrix}
	\right) = \mathcal{S}_{\rm in} \left(\begin{matrix}
	a^e_L\\[.25em]a^e_R\\[.25em]a^h_L\\[.25em]a^h_R
	\end{matrix}
	\right),
	\end{equation}
	we can model the perfect Andreev reflections at $x=\pm L/2$ with the matrix $\mathcal{S}_{\rm And}$ which reads
	\begin{equation}
	\begin{split}
	\label{eq:SAnd}
	&\left(\begin{matrix}
	a^e_L\\[.25em]a^e_R\\[.25em]a^h_L\\[.25em]a^h_R
	\end{matrix}
	\right)
	= \mathcal{S}_{\rm And}
	\left(\begin{matrix}
	b^e_L\\[.25em]b^e_R\\[.25em]b^h_L\\[.25em]b^h_R
	\end{matrix}
	\right)
	= e^{i L \epsilon} e^{-i \arccos\left(\frac{\epsilon}{\Delta_0}\right)}\; \times\\ &\quad \times
	\left(\begin{matrix}
	0&0&e^{-i(\chi_L+L\tilde m_z)}&0\\
	0&0&0&e^{-i(\chi_R-L\tilde m_z)}\\
	e^{i(\chi_L+L\tilde m_z)}&0&0&0\\
	0&e^{i(\chi_R-L\tilde m_z)}&0&0
	\end{matrix}
	\right)
	\left(\begin{matrix}
	b^e_L\\[.25em]b^e_R\\[.25em]b^h_L\\[.25em]b^h_R
	\end{matrix}
	\right).
	\end{split}
	\end{equation}
	Taking into account the scattering in the inner gray region as well as the Andreev reflections at $x=\pm L/2$, one can derive the well-known compatibility condition \cite{Crepin2014} for the existence of bound states with energy $\epsilon$
	\begin{equation}
	\det\left[1-\mathcal{S}_{\rm And}(\epsilon)\mathcal{S}_{\rm in}(\epsilon)\right]  =0.
	\end{equation}
	Note that the presence of magnetization along $z$ in the gapless region $\tilde m_z\neq 0$ is harmless since it only corresponds to a shift of the two superconducting phases.

	\section{Effects of finite chemical potential and/or perpendicular magnetization}
	\label{app:C}
\begin{figure}
		\centering
		\includegraphics[width=0.9\linewidth]{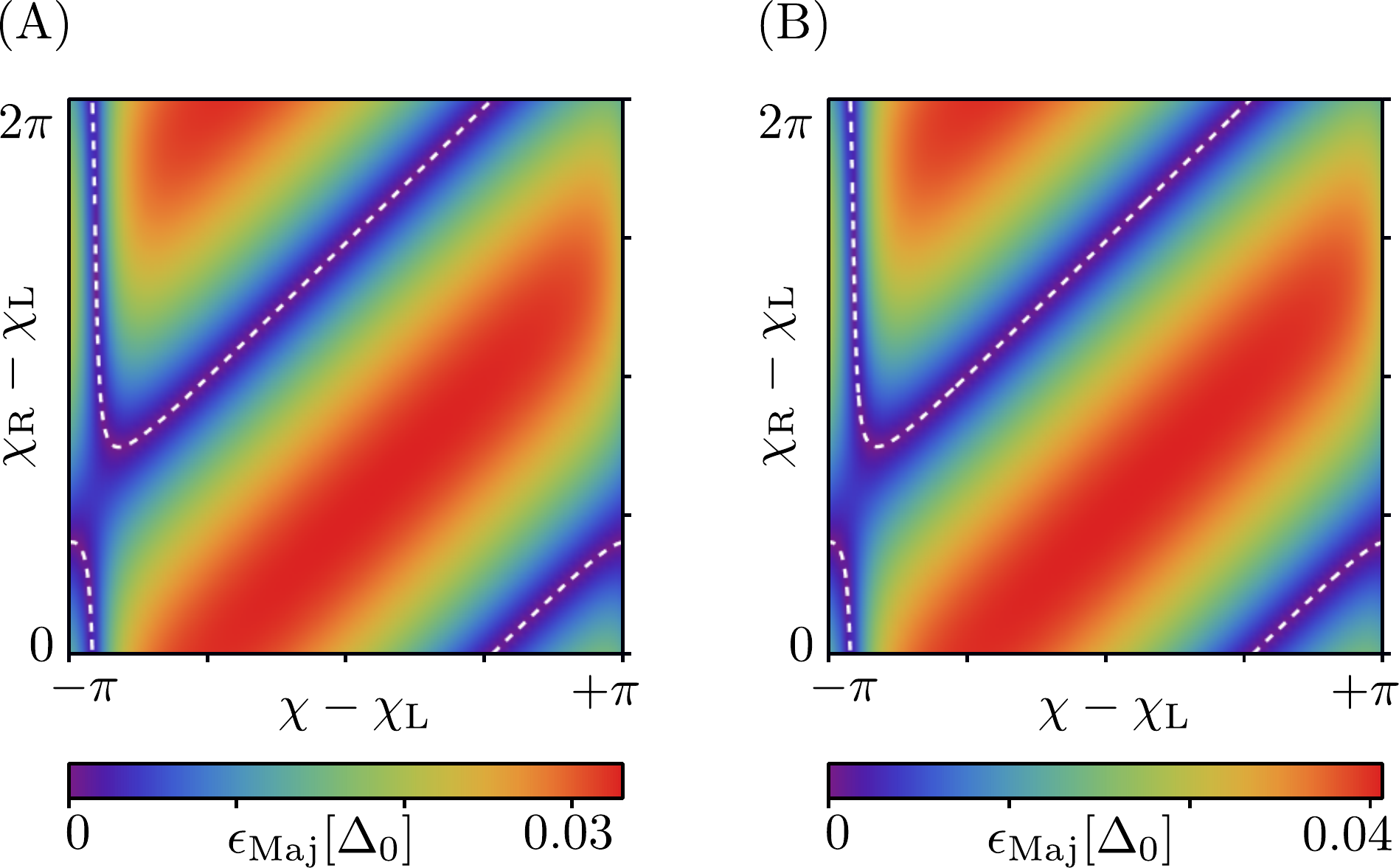}
		\caption{Energy splitting $\epsilon_{\rm Maj}$ (units $\Delta_0$) of the two Majoranas as a function of the superconducting phase differences. The white dashed line corresponds to the condition in Eq.~\eqref{eq:chiR_mu} and nicely highlights the presence of zero-energy MBS. In the left panel, the ferromagnet features a non-vanishing perpendicular magnetization $m_z=0.2\,\Delta_0$ while all the gapless regions feature $\tilde m_z = 0.15 \,\Delta_0$. In the right panel, we add non vanishing chemical potentials: $\mu_S=4\,\Delta_0$, $\mu_F=0.5\,\Delta_0$, $\tilde \mu=0.3\,\Delta_0$ within the superconductor, the ferromagnet and the gapless regions, respectively. The behavior of the system at zero-energy is not affected by the chemical potential.  As for the remaining parameters, in analogy with Fig.~\ref{fig:fig2} (A) of the main text, we choose $\Delta=m=\Delta_0$ and $l_{\rm S}=l_{\rm F}=\xi=\Delta_0^{-1}$,$L=7\xi$, $x_{\rm F}=-x_{\rm S}=1.5 \xi$.}
		\label{fig:smfig3}
	\end{figure}

	For the sake of simplicity, in the main text we focus only on configurations with vanishing chemical potential and zero magnetization along $z$. Importantly, we argue that their presence does not qualitatively modify the zero-energy physics of the system. To this end, it is useful to understand how Eqs. (\ref{eq:Scat2}-\ref{eq:rho}) of the main text are modified. In the presence of non-vanishing chemical potential and/or magnetization along $z$, the finite superconductor acts as
	\begin{widetext}
	\begin{equation}
	\left(\begin{matrix}
	b^h_{\rm L} \\[.25em]b^e_{\rm R}
	\end{matrix}\right)  = \left(\begin{matrix}
	-i e^{i \chi} \exp\left[i \tilde m_z (l-2x_0)\right] \tanh  (\Delta\, l_{\rm S}) & \exp\left[il(\tilde m_z-\mu+\tilde \mu)\right]\;\text{sech} (\Delta\, l_{\rm S})\\[.25em]
	\exp\left[il(\tilde m_z+\mu-\tilde \mu)\right]\; \text{sech} (\Delta\, l_{\rm S}) & 	-i e^{-i \chi}  \exp\left[i \tilde m_z (l+2x_0)\right] \tanh  (\Delta\, l_{\rm S})
	\end{matrix}\right)\left(\begin{matrix}
	a^e_{\rm L} \\[.25em] a^h_{\rm R}
	\end{matrix}\right).
	\end{equation}
	The perfect Andreev reflection at $x=-L/2$ relates
	\begin{equation}
	a^e_{\rm L} = -i e^{-i\chi_{\rm L}-L \tilde m_z} b^h_{\rm L}.
	\end{equation}
	The combined effect of the two superconductors leads to the relation $b^e_{\rm R} = -i e^{-i \tilde \chi_{\rm eff}} a^h_{\rm R}$, with 
	\begin{equation}
	\begin{split}
	\tilde\chi_{\rm eff} (\chi_L,\chi,\Delta\,  l_{\rm S})&= \chi - \arg \left[\frac{\exp[i\chi + i2l_{\rm S}\tilde m_z]+\exp[i\chi_L + i\tilde m_z (L+l_{\rm S}+2x_0)] \tanh(\Delta \, l_{\rm S})}
	{\exp[i \chi_L+i\tilde m_z L]+\exp  [i\chi + i\tilde m_z (l_{\rm S}-2x_0)] \tanh(\Delta \, l_{\rm S})}\right]\\
	&=\chi - (l_{\rm S}+2x_0)\tilde m_z - \arg \left[\frac{\exp[i\chi + i\tilde m_z(l_{\rm S}-2x_0-L)]+\exp[i\chi_L ] \tanh(\Delta \, l_{\rm S})}
	{\exp[i \chi_L]+\exp  [i\chi + i\tilde m_z (l_{\rm S}-2x_0-L)] \tanh(\Delta \, l_{\rm S})}\right]\\
	\end{split}
	\end{equation}
\end{widetext}
	Using the known result for the SFS geometry \cite{Crepin2014}, we conclude that the system host zero-energy modes whenever
	\begin{equation}
	\label{eq:chiR_mu}
	\chi_R  = \tilde\chi_{\rm eff} + L \tilde m_z + 2 l_{\rm F} (m_z-\tilde m_z)+\pi.
	\end{equation}
	The validity of this relation is nicely verified in Fig.~\ref{fig:smfig3} where we plot the energy splitting $\epsilon_{\rm Maj}$ as a function of the phase differences $\chi_{\rm R}-\chi_{\rm L}$ and $\chi-\chi_{\rm L}$, in presence of finite chemical potentials and magnetizations along $z$. As for the localization of the Majoranas, we find
	\begin{equation}\begin{split}
	\rho&= \left|\frac{b_e}{a_e}\right|^2 =  \cosh(2\Delta\,l_{\rm S})\\&\qquad + \sinh(2\Delta\, l_{\rm S}) \cos(\chi_{\rm L}-\chi-\tilde m_z(l_{\rm S}-L-2x_0)).
	\end{split}
	\end{equation}
	We observe that both the localization of Majoranas and the condition to have zero-energy modes depends on the parameter $\chi-\chi_L+\tilde m_z(l_S-L-2x_0)$. 
	
	We can, therefore, conclude that the presence of finite chemical potentials and/or magnetizations along the $z$ axis do not significantly affect the behavior of the zero-energy MBS. Indeed, the latter have proved to be insensitive to variations of the chemical potential while the presence of non-vanishing magnetization along $z$, both in the ferromagnetic and gapless regions, merely corresponds to shifts in the superconducting phases. This particular behavior, which is related to the topological origin of the MBS in our system \cite{Crepin2014}, justifies to safely consider only the $\tilde m_z=\tilde \mu=\mu=0$ case in the main text. 
	
	\section{The quantum point contact}
	\label{app:D}
	Let us now consider a QPC between two helical edges with opposite helicity. We choose the the upper one to have ``$+$'' helicity, i.e. right-movers electrons have spin-up, and the lower one to have ``$-$'' helicity. At first, we focus on two gapless edges, without considering ferromagnets or superconductors. The free Hamiltonian simply reads
	\begin{equation}
	H_{0} = \sum_{\sigma=\uparrow,\downarrow}\sum_{r=R,L} \int dx\; \vartheta_r \; \psi^\dagger_{r\sigma}(x)(-i\partial_x) \psi_{r\sigma}(x)
	\end{equation}
	with $\vartheta_{R/L}=\pm1$. A QPC located at $x=\bar x$ is described by the Hamiltonian
	\begin{equation}\begin{split}
	H_{\rm QPC} &= \sum_{\sigma=\uparrow,\downarrow} \left(2\lambda_{sp} \; \psi^\dagger_{R\sigma}(\bar x) \psi_{L\sigma}(\bar x)\right) \\&\quad + \sum_{r=R,L} \left(2\vartheta_r \lambda_{sf} \; \psi^\dagger_{r\uparrow}(\bar x) \psi_{r\downarrow}(\bar x)\right) + h.c.
	\end{split}\end{equation}
	where $\lambda_{sp}$ ($\lambda_{sf}$) is the spin-preserving (spin-flipping) tunneling amplitude \cite{Ferraro2014}. The Heisenberg equation of motion
	\begin{equation}
	i\partial_t \psi_{r\sigma} = \left[\psi_{r\sigma},H_{\rm QPC}+H_0]\right]
	\end{equation}
	gives us the following set of differential equations
	\begin{align}
	i\left(\partial_t + \partial_x \right)\psi_{R\uparrow} &= 2\delta(x-\bar x) \left[ \lambda_{sp} \psi_{L\uparrow} + \lambda_{sf} \psi_{R\downarrow}\right],
	\\
	i\left(\partial_t + \partial_x \right)\psi_{R\downarrow} &= 2\delta(x-\bar x)
	\left[ \lambda_{sp} \psi_{L\downarrow} + \lambda_{sf}^* \psi_{R\uparrow}\right],
	\\
	i\left(\partial_t - \partial_x \right)\psi_{L\uparrow} &= 2\delta(x-\bar x)
	\left[ \lambda_{sp}^* \psi_{R\uparrow} - \lambda_{sf} \psi_{L\downarrow}\right],
	\\
	i\left(\partial_t - \partial_x \right)\psi_{L\downarrow} &= 2\delta(x-\bar x)
	\left[ \lambda_{sp}^* \psi_{R\downarrow} - \lambda_{sf}^* \psi_{L\uparrow}\right].
	\end{align}
	The plane-wave ansatz allows us to solve these equations of motions (by integrating them over an infinitesimal interval around $x=\bar x$). The incoming ($\alpha_{r\sigma}$) and outgoing ($\beta_{r\sigma}$) amplitudes satisfy the following set of equations
	\begin{align}
&	i(\beta_{R\uparrow}-\alpha_{R\uparrow}) = \lambda_{sp}(\alpha_{L\uparrow}+\beta_{L\uparrow}) \exp(-2i\epsilon \bar x) \notag\\& \qquad +  \lambda_{sf}(\alpha_{R\downarrow}+\beta_{R\downarrow}),\\[.75em]
&	i(\beta_{R\downarrow}-\alpha_{R\downarrow}) = \lambda_{sp}(\alpha_{L\downarrow}+\beta_{L\downarrow})\exp(-2i\epsilon \bar x)\notag\\&  \qquad+  \lambda_{sf}^*(\alpha_{R\uparrow}+\beta_{R\uparrow}),\\[.75em]
&	\exp(-2i\epsilon \bar x) \; i(\beta_{L\uparrow}-\alpha_{L\uparrow}) = \lambda_{sp}^*(\alpha_{R\uparrow}+\beta_{R\uparrow}) \notag\\&  \qquad- \lambda_{sf}(\alpha_{L\downarrow}+\beta_{L\downarrow})\exp(-2i\epsilon \bar x),\\[.75em]
&	\exp(-2i\epsilon \bar x) \; i(\beta_{L\downarrow}-\alpha_{L\downarrow}) = \lambda_{sp}^*(\alpha_{R\downarrow}+\beta_{R\downarrow})\notag\\&  \qquad- \lambda_{sf}^*(\alpha_{L\uparrow}+\beta_{L\uparrow})\exp(-2i\epsilon \bar x),
	\end{align}
	which can be recast as
	\begin{equation}
	\left(\begin{matrix}
	b_L^{e[1]}\\[.25em]
	b_R^{e[1]}\\[.25em]
	b_L^{e[2]}\\[.25em]
	b_R^{e[2]}\end{matrix}\right)=
	\left(\begin{matrix}
	\beta_{L\downarrow}\\[.25em]
	\beta_{R\uparrow}\\[.25em]
	\beta_{L\uparrow}\\[.25em]
	\beta_{R\downarrow}
	\end{matrix}\right) =  \mathcal{S}_{\rm QPC}^{(e)} \left(\begin{matrix}
	a_L^{e[1]}\\[.25em]
	a_R^{e[1]}\\[.25em]
	a_L^{e[2]}\\[.25em]
	a_R^{e[2]}
	\end{matrix}\right) =  \mathcal{S}_{\rm QPC}^{(e)} \left(\begin{matrix}
	\alpha_{R\uparrow}\\[.25em]
	\alpha_{L\downarrow}\\[.25em]
	\alpha_{R\downarrow}\\[.25em]
	\alpha_{L\uparrow}
	\end{matrix}\right)
	\end{equation}
	where the scattering matrix associated with the QPC reads
	\begin{equation}
	\label{eq:SqpcE}
	\mathcal{S}_{\rm QPC}^{(e)} =  \left( \begin{matrix}
	0	&	\Lambda_{pf}& -\Lambda_{pb}^* & -\Lambda_{ff}^* \\
	\Lambda_{pf}& 0 &  -\Lambda_{ff} & 	\Lambda_{pb}\\
	-\Lambda_{pb}^*	&	\Lambda_{ff} & 0 &	\Lambda_{pf} \\
	\Lambda_{ff}^*	& \Lambda_{pb} &	\Lambda_{pf}&  0 \\
	\end{matrix}\right)
	\end{equation}
	and
	\begin{align}
	\Lambda_{pb} &=  \exp(-2i\epsilon \bar x) \frac{-2i \lambda_{sp}}{1+|\lambda_{sp}|^2+|\lambda_{sf}|^2}\\
	\Lambda_{ff} &=   \frac{2i \lambda_{sf}}{1+|\lambda_{sp}|^2+|\lambda_{sf}|^2}\\
	\Lambda_{pf} &=   \frac{1-|\lambda_{sp}|^2-|\lambda_{sf}|^2}{1+|\lambda_{sp}|^2+|\lambda_{sf}|^2}
	\end{align}
	A remark on the notation: $a_L^{e[1]}$ is the incoming amplitude ($a$) from the left ($L$) of the QPC for electrons ($e$) on the upper edge ($1$); the same applies for the other amplitudes.\\ 
	
	\subsection{Scattering matrix in Nambu space}
	
	In order to take into account the presence of superconducting and ferromagnetic gapped regions on the edges, it is useful to express the QPC scattering matrix taking also into account tunneling of holes. We thus introduce $8$-dimensional vectors of incoming and outgoing scattering amplitudes, $\vec{a}_8$ and $\vec{b}_8$ respectively
	\begin{equation}
	\vec{a}_8 = \left(\begin{matrix}
	a^{e[1]}_L\\[.25em]a^{e[1]}_R\\[.25em]a^{h[1]}_L\\[.25em]a^{h[1]}_R\\a^{e[2]}_L\\[.25em]a^{e[2]}_R\\[.25em]a^{h[2]}_L\\[.25em]a^{h[2]}_R
	\end{matrix}
	\right),\qquad
	\vec{b}_8 =\left(\begin{matrix}
	b^{e[1]}_L\\[.25em]b^{e[1]}_R\\[.25em]b^{h[1]}_L\\[.25em]b^{h[1]}_R\\b^{e[2]}_L\\[.25em]b^{e[2]}_R\\[.25em]b^{h[2]}_L\\[.25em]b^{h[2]}_R
	\end{matrix}
	\right)
	%,\qquad
	%\vec{l}_8= \left(\begin{matrix}
	%a^{e1}_L\\[.25em]b^{e1}_L\\[.25em]a^{h1}_L\\[.25em]b^{h1}_L\\a^{e2}_L\\[.25em]b^{e2}_L\\[.25em]a^{h2}_L\\[.25em]b^{h2}_L
	%\end{matrix}
	%\right),\qquad
	%\vec{r}_8 = \left(\begin{matrix}
	%b^{e1}_R\\[.25em]a^{e1}_R\\[.25em]b^{h1}_R\\[.25em]a^{h1}_R\\b^{e2}_R\\[.25em]a^{e2}_R\\[.25em]b^{h2}_R\\[.25em]a^{h2}_R
	%\end{matrix}
	%\right).
	\end{equation}
	By exploiting the particle-hole symmetry of the system, we can obtain the expression for the $8$-dimensional QPC scattering matrix
	\begin{widetext}
	\begin{equation}\begin{split}
	&\mathcal{S}_{\rm QPC}
	%(\lambda_{sf},\lambda_{sp},\epsilon,\bar x) =\\&\qquad
	= \left(\begin{matrix}
	0&	\Lambda_{pf}&&&-\Lambda_{pb}^*(\epsilon) & - \Lambda_{ff}^*&&\\[.25em]
	\Lambda_{pf}&0&&&- 	\Lambda_{ff} & 	\Lambda_{pb}(\epsilon)&&\\[.25em]
	&&0	&	\Lambda_{pf}&&&\Lambda_{pb}(-\epsilon) &  -\Lambda_{ff}\\[.25em]
	&&	\Lambda_{pf}&0&&& 	-\Lambda_{ff}^* & 	-\Lambda_{pb}^*(-\epsilon)\\[.25em]
	-\Lambda_{pb}^*(\epsilon) &  \Lambda_{ff}&&&0&	\Lambda_{pf}&&\\[.25em]
	\Lambda_{ff}^* & 	\Lambda_{pb}(\epsilon)&&&	\Lambda_{pf}&0&&\\[.25em]
	&&\Lambda_{pb}(-\epsilon) &  \Lambda_{ff}^*&&&0&	\Lambda_{pf}\\[.25em]
	&&	\Lambda_{ff} & 	-\Lambda_{pb}^*(-\epsilon)&&&	\Lambda_{pf}&0\\
	\end{matrix}
	\right)
	\end{split}
	\end{equation}
\end{widetext}
	which relates $\vec{b}_8= \mathcal{S}_{\rm QPC}\; \vec{a}_8$. Note that the $8$-dimensional scattering matrix associated with regions where no inter-edge processes are present is block diagonal. % For example, the effect of a gapped region on the upper edge (described by the single-edge scattering matrix $\mathcal{S}_4$) is captured by the scattering matrix
	%\begin{equation}\begin{split}
	%&\mathcal{S}_{8}
	%(\lambda_{sf},\lambda_{sp},\epsilon,\bar x) =\\&\qquad
	%= \left(\begin{matrix}
	%\mathcal{S}_4 & 0\\
	%0& \text{I}_4\\
	%\end{matrix}
	%\right)
	%\end{split}
	%\end{equation}
	%where
	%\begin{equation}
	%\text{I}_4 = \left(
	%\right)
	%\end{equation}
	%is a scattering matrix with unitary transmissions. 

	\subsection{Coupling MBS with a QPC}
	We now address in more detail the effect of electron tunneling at the QPC. While it definitely induces some coupling between the two edges, it is not obvious that it can lead to the hybridization of two zero-energy MBS located on different edges. Indeed, given two generic Majorana operators $\gamma$ and $\eta$, they can only be coupled by an imaginary hopping amplitude $H_{\rm hop} =  i\;  \eta\gamma$. As a result, one has to carefully implement phase differences between the two edges. 
	
			\begin{figure}
		\centering
		\includegraphics[width=.95\linewidth]{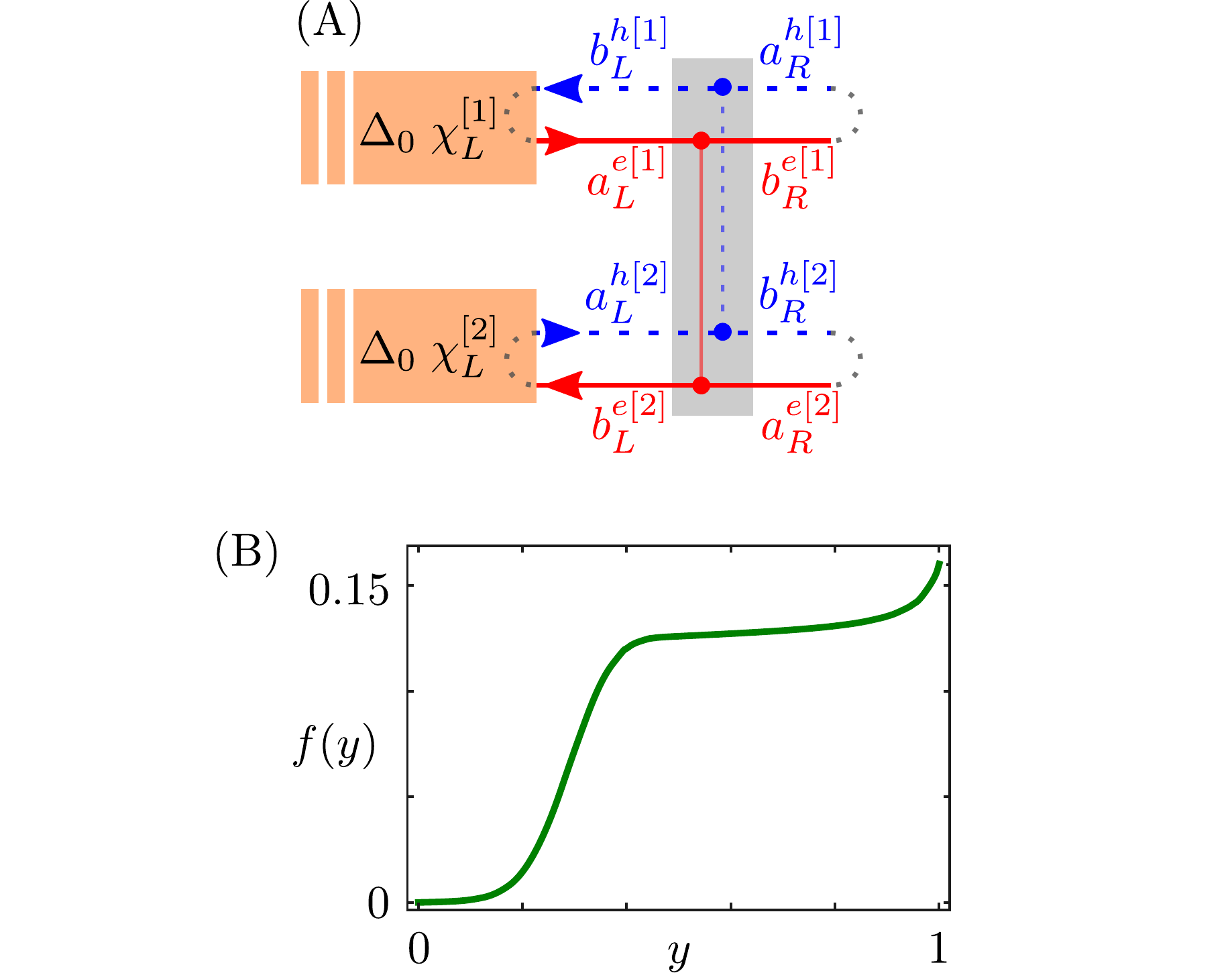}
		\caption{(A) Sketch of spin-preserving tunnelings between the two edges. We focus only on the left-most gapless regions of the system, depicted in Fig.~\ref{fig:DE} of the main text. Red (blue) lines refer to spin-up electron (spin-down hole) channels. (B) Plot of the correction phase $f(y)$ (in units of $\pi$). As in the main text, we choose $m^{[1]}=2m^{[2]}$, $\Delta=m^{[2]}=\Delta_0$.}
		\label{fig:smfig2}
	\end{figure}

	In order to develop some intuition, it is useful to focus on the simple example depicted in Fig.~\ref{fig:smfig2} (A). We focus only on the left-most gapless regions of the two edges, assuming that they host zero-energy MBS. We study the effect of a weak spin-preserving tunneling of electrons (holes) with spin up (down). The existence of MBS at zero energy requires
	\begin{equation}
	\left(\begin{matrix}
	a_L^{e[1]} \\[.2em]a_R^{h[1]}\\[.2em]a_R^{e[2]} \\[.2em]a_L^{h[2]}
	\end{matrix}\right) = -i \, \left(\begin{matrix}
	e^{- i\chi_L^{[1]}} & 0&0&0\\[.2em]
	0& e^{ i\chi_L^{[1]}+i\pi}&0&0\\[.2em]
	
	0&0&e^{ i\chi_L^{[2]}} & 0\\[.2em]
	0&0&0& e^{- i\chi_L^{[2]}+i\pi}
	\end{matrix}\right)\left(\begin{matrix}
	b_L^{h[1]} \\[.2em]b_R^{e[1]}\\[.2em]b_R^{h[2]} \\[.2em]b_L^{e[2]}
	\end{matrix}\right).
	\end{equation}
	The spin-preserving tunneling at zero-energy is described by the (reduced) scattering matrix 
	\begin{equation}
	\left(\begin{matrix}
	b_L^{h[1]} \\[.2em]b_R^{e[1]}\\[.2em]b_R^{h[2]} \\[.2em]b_L^{e[2]}
	\end{matrix}\right) = \mathcal{\bar S}_{\rm QPC} 
	\left(\begin{matrix}
	a_L^{e[1]} \\[.2em]a_R^{h[1]}\\[.2em]a_R^{e[2]} \\[.2em]a_L^{h[2]}
	\end{matrix}\right)
	=  
	\left(\begin{matrix}
	0&\Lambda_{\rm f}&0&\Lambda_{\rm b}\\[.2em]
	\Lambda_{\rm f}&0&\Lambda_{\rm b}&0\\[.2em]
	0&-\Lambda_{\rm b}^*&0&\Lambda_{\rm f}\\[.2em]
	-\Lambda_{\rm b}^*&0&\Lambda_{\rm f}&0
	\end{matrix}\right)
	\left(\begin{matrix}
	a_L^{e[1]} \\[.2em]a_R^{h[1]}\\[.2em]a_R^{e[2]} \\[.2em]a_L^{h[2]}
	\end{matrix}\right)
	\end{equation}
	with forward (f) and backscattering (b) entries
	\begin{equation}
	\Lambda_{\rm f} = \frac{1-|\lambda_{sp}|^2}{1+|\lambda_{sp}|^2},\quad \Lambda_{\rm b} = \frac{-2i \lambda_{sp}}{1+|\lambda_{sp}|^2}.
	\end{equation}
	We can now inspect the effects of the QPC %at the lowest order ($|\lambda_{sp}|^4$) 
	by considering the fate of a right moving electron with spin up, which has just emerged from the left superconductor on the upper edge. It can follows two trajectories: it can remain on the upper edge being Andreev reflected (i); or it can tunnel on the other edge, being Andreev reflected by the left superconductor, and tunnel back (ii). Depending on the trajectory, it eventually reemerges as an electron on the upper edge with corresponding prefactors
	\begin{align}
	\xi_{(i)} &= \Lambda_{\rm f} (-i e^{i\chi_L^{[1]}+i\pi} ) \Lambda_{\rm f} (-i e^{-i\chi_L^{[1]}} ) = 1-|\Lambda_{\rm b}|^2,
	\\
	\xi_{(ii)} &= -\Lambda_{\rm p}^* (-i e^{i\chi_L^{[2]}}) \Lambda_{\rm p} (-i e^{-i\chi_L^{[1]}} ) = e^{i (\chi_L^{[1]} - \chi_L^{[2]})} |\Lambda_{\rm b}|^2.
	\end{align}
	This clearly shows that, if no phase difference between the edges is considered $\chi_L^{[1]}-\chi_L^{[2]}=0$, spin-preserving tunneling of electrons is ineffective in coupling zero-energy MBS. 
	
	\section{Local density of states}
	\label{app:E}
	Our goal is to compute the equilibrium zero-temperature local density of states (lDOS) on a given edge. This observable is defined as 
	\begin{equation}
	\begin{split}
	&\mathcal{A}(x,\omega)=-\frac{1}{\pi} \text{Im}\left[\text{Tr }G^R_{ee}(x,x,\omega)]\right]\\
	&= -\frac{1}{\pi}  \sum_{\sigma=\uparrow,\downarrow} \text{Im} \int d\tau \; e^{i\omega \tau - 0^+\tau}\;  \theta(\tau) \left[-i \left\langle
	\left\{\psi_{\sigma}(x,t),\, \psi^\dagger_\sigma(x,t-\tau)\right\}
	\right\rangle_0\right]\\
	&= \frac{1}{2\pi}  \sum_{\sigma=\uparrow,\downarrow} \int d\tau \; e^{i\omega \tau - 0^+|\tau|}\; \left\langle
	\left\{\psi_{\sigma}(x,t),\, \psi^\dagger_\sigma(x,t-\tau)\right\}
	\right\rangle_0
	\end{split}
	\end{equation}
	where $\langle \dots \rangle_0$ is the average with respect to the groundstate. By using Eqs.~\eqref{eq:Nambu} and \eqref{eq:ansatz}, the fermionic fields $\psi_\sigma(x)$ can be expressed as ($\sigma=\, \uparrow,\downarrow\, =\pm$)
	\begin{equation}
	\psi_\sigma(x,t) = \sum_{\epsilon} \sum_j \left[ u_{\sigma,j} (\epsilon,x) \gamma_{\epsilon,j}(t) - \sigma v_{\sigma,j}(\epsilon,x)^* \gamma_{\epsilon,j}^\dagger(t)\right].
	\end{equation}
	Given the system Hamiltonian \eqref{eq:BdGdiag}, it is straightforward to compute the groundstate averages in the non-degenerate case
	\begin{equation}
	\begin{split}
	&\left\langle
	\left\{\psi_{\sigma}(x,t),\, \psi^\dagger_\sigma(x,t-\tau)\right\}
	\right\rangle_0 =\\&= \sum_{\epsilon,\epsilon'\geq 0} 
	\left(	u_\sigma(\epsilon,x)u_\sigma(\epsilon',x)^* e^{-i\epsilon t+i\epsilon'(t-\tau)} \right.\\&\qquad+\left. v_\sigma(\epsilon,x)v_\sigma(\epsilon',x)^* e^{+i\epsilon t-i\epsilon'(t-\tau)}\right)
	\left\langle \gamma_{\epsilon}\gamma_{\epsilon'}^\dagger	\right\rangle_0\\
	&= \sum_{\epsilon\geq 0} \left( |u_\sigma(\epsilon,x)|^2 e^{-i\epsilon\tau}+|v_\sigma(\epsilon,x)|^2e^{+i\epsilon\tau}\right).
	\end{split}
	\end{equation}
	The lDOS can be expressed as a sum of $\delta$-functions centered on the energies of the bound states. By exploiting Eq.~\eqref{eq:uv_epsilon}, we get
	\begin{equation}
	\begin{split}
	\label{eq:Anondeg}
	\mathcal{A}(x,\omega) & =  \sum_{\epsilon} \sum_{\sigma} |u_\sigma(\epsilon,x)|^2  \delta(\omega+\epsilon)= \sum_{\epsilon}  A_\epsilon(x)\;  \delta(\omega+\epsilon)
	\end{split}
	\end{equation}
	In presence of zero-energy Majoranas, the groundstate is degenerate. In order to deal with this subtlety, we numerically introduce a tiny perturbation in one of the phases, so that the degeneracy is lifted by a very small amount ($\zeta\lesssim 10^{-8}\Delta_0$) and we computed $|u_\sigma(\pm\zeta,x)|^2$. The zero-energy lDOS plotted in the main text is then given by
	\begin{equation}
	A_0(x) \simeq \sum_{\sigma} |u_\sigma(\zeta,x)|^2 + |u_\sigma(-\zeta,x)|^2
	\end{equation}
	
	%\subsection{Degeneracies}
	%So far, we haven't discussed the presence of degenereacies. While it is straightforward to take into account the existence of degenerate boundstates at finite energy, it is important to carefully discuss the case with degenerate zero-energy boundstates. In particular, because of the particle-hole symmetry of the system, boundstates at zero energy always come in pair. Furthermore, they can be expressed in terms of Majorana operators \cite{Crepin2014}. 
	
	%Let us assume that the system hosts two boundstates at zero energy $\varphi_{0,j=A}(x)$ and $\varphi_{0,j=B}(x)=\mathcal{C} \left[ \varphi_{0,j=A}(x)\right]$. One can show that \cite{Crepin2014}
	%\begin{equation}
	%\gamma_{0,A}^\dagger = \gamma_{0,B}
	%\end{equation} 
	
	\section{Existence of zero-energy modes}
	\label{app:F}
	\subsection{A generic result}
	We first discuss a generic property of helical systems connected to semi-infinite superconductors. In particular, we consider one semi-infinite superconductor located at $x\geq L/2$, i.e. to the right of the helical system (see Fig.~\ref{fig:smfig1}). Moreover, we make two assumptions regarding what happens to the left of the gapless helical system: (i) all the amplitudes going to the left ($a^h_R$ and $a^e_R$) are completely reflected back, i.e. the system obeys some kind of open boundary conditions; (ii) the whole system is particle-hole symmetric. Our goal is to argue that it is always possible to tune the phase $\chi_R$ of the right semi-infinite superconductor in order for the whole system to support zero-energy bound states. \\
	
	The first assumption assures that the matrix $U(\epsilon)$ 
	\begin{equation}
	\left(\begin{matrix}
	b^e_R(\epsilon)\\b^h_R(\epsilon)
	\end{matrix}\right) = 
	U(\epsilon)
	\left(\begin{matrix}
	a^h_R(\epsilon)\\a^e_R(\epsilon)
	\end{matrix}\right)
	\end{equation}
	is unitary. The second one, by using Eq.~\eqref{eq:uv_epsilon}, leads to 
	\begin{equation}
	\begin{split}	&-\left(\begin{matrix}
	b^h_R(-\epsilon)^*\\b^e_R(-\epsilon)^*
	\end{matrix}\right) = 
	U(\epsilon)
	\left(\begin{matrix}
	a^e_R(-\epsilon)^*\\a^h_R(-\epsilon)^*
	\end{matrix}\right)
	\\&\qquad \Rightarrow
	-\left(\begin{matrix}
	b^h_R(\epsilon)\\b^e_R(\epsilon)
	\end{matrix}\right) = 
	-U(-\epsilon)^*
	\left(\begin{matrix}
	a^e_R(\epsilon)\\a^h_R(\epsilon).
	\end{matrix}\right)
	\end{split}
	\end{equation}
	At zero energy, the unitary matrix $U(0)$ therefore obeys
	\begin{equation}
	\begin{split}
	U(0)&=e^{i\varphi}\left(
	\begin{matrix}
	\sin(\theta) e^{i\alpha} & \cos(\theta) e^{i\beta} \\
	-\cos(\theta) e^{-i\beta} & \sin(\theta) e^{-i\alpha} 
	\end{matrix}
	\right)\\&=-\left(\begin{matrix}
	0&1\\1&0
	\end{matrix}\right)U(0)^*\left(\begin{matrix}
	0&1\\1&0
	\end{matrix}\right)
	\end{split}
	\end{equation}
	which poses constraints on the parameters $\alpha,\beta,\theta$ and $\varphi$. In particular, there are only two possibilities: either $\theta=0$ and $\varphi=0,\pi$ or $\theta=\pi/2$ and $\varphi=\pi/2, 3\pi/2$. As for the right semi-infinite superconductor, at zero energy, it Andreev reflects the incoming wavefunction according to  Eq.~\eqref{eq:SAnd}
	\begin{equation}
	\left(\begin{matrix}
	a^h_R\\a^e_R
	\end{matrix}\right)=-i
	\left(\begin{matrix}
	e^{i\chi_R}&0\\
	0& e^{-i\chi_R}
	\end{matrix}\right)\left(\begin{matrix}
	b^e_R\\b^h_R
	\end{matrix}\right).
	\end{equation}
	The whole system admits bound states at zero-energy if it is possible to solve
	\begin{equation}
	\begin{split}
	\left(\begin{matrix}
	b^e_R\\b^h_R
	\end{matrix}\right) &= 
	e^{i\varphi}\left(
	\begin{matrix}
	\sin(\theta) e^{i\alpha} & \cos(\theta) e^{i\beta} \\
	-\cos(\theta) e^{-i\beta} & \sin(\theta) e^{-i\alpha} 
	\end{matrix}
	\right)\left(\begin{matrix}
	-i e^{i\chi_R}&0\\
	0&-i e^{-i\chi_R}
	\end{matrix}\right)\left(\begin{matrix}
	b^e_R\\b^h_R
	\end{matrix}\right) \\&=e^{i\varphi-i\tfrac{\pi}{2}}\left(
	\begin{matrix}
	\sin(\theta) e^{i\alpha}e^{i\chi_R} & \cos(\theta) e^{i\beta}e^{-i\chi_R} \\
	-\cos(\theta) e^{-i\beta}e^{i\chi_R} & \sin(\theta) e^{-i\alpha} e^{-i\chi_R}
	\end{matrix}
	\right)
	\left(\begin{matrix}
	b^e_R\\b^h_R
	\end{matrix}\right)
		\end{split}\end{equation}
	which is equivalent to require
	\begin{equation}
	\begin{split}
	&\det \left[1-e^{i\varphi-i\tfrac{\pi}{2}}\left(
	\begin{matrix}
	\sin(\theta) e^{i\alpha}e^{i\chi_R} & \cos(\theta) e^{i\beta}e^{-i\chi_R} \\
	-\cos(\theta) e^{-i\beta}e^{i\chi_R} & \sin(\theta) e^{-i\alpha} e^{-i\chi_R}
	\end{matrix}
	\right) \right] =0 \\
	& \Rightarrow \sin(\theta)\cos(\alpha+\chi_{\rm R})-\sin(\varphi) = 0
	\end{split}
	\end{equation}
	In the $\theta=0$ case, there is therefore always a solution at zero energy regardless of the values of $\beta$ and $\chi_R$. In the $\theta=\pi/2$ case, it is possible to tune $\chi_R=\varphi-\pi/2-\alpha$ which results in zero-energy bound states.

	\subsection{Zero-energy modes in the protocol discussed in the main text}
	We are now in a position to comment on the tiny correction $f(y)$ which appears in the protocol that moves Majoranas from one edge to the other one. Without implementing such a correction, the protocol turns out to be qualitatively correct but it fails in keeping the MBS exactly at zero-energy. This has to be expected since the protocol has been designed by exploiting the exact knowledge of the single-edge SSFS architectures supplemented with an approximate and qualitative description of the QPC.
	
	Interestingly, it turns out that this subtlety can be straightforwardly fixed. The general result discussed above ensures that, for each value of $y$, one can force the MBS to be at zero energy just by adding a proper correction $f(y)$ to one phase, say $\chi_R^{[1]}$. The function $f(y)$ can be easily computed by numerically requiring the MBS to be at zero energy and the result is plotted in Fig.~\ref{fig:smfig2} (right panel). Remarkably enough, it turns out to be a smooth function which features only small values ($0\leq f(y) \leq 0.15 \pi$) and therefore does not qualitatively affect the protocol.  

	\bibliography{../../../Refs/Ref,notes.bib}

\begin{thebibliography}{61}
\bibitem [{not()}]{note1}%
  \BibitemOpen
  \href@noop {} {}\bibinfo {note} {In the limiting case $(\Delta l_S) \to
  \infty$, the system effectively consist of two uncoupled junctions, SS and
  SFS, which can both accommodate a couple of zero-energy Majoranas. When the
  central superconductor has finite strength $(\Delta l_S)$, there is a finite
  coupling between the junctions proportional to $\exp[-\Delta l_S]$ and the
  four Majoranas hybridize. As a result, only a single pair of Majoranas can be
  at zero energy while the other pair will always feature a minimal
  hybridization energy proportional to $\exp[-\Delta l_S]$}\BibitemShut
  {NoStop}%
\end{thebibliography}
	
\end{document}